# A digital-twin framework for forecasting treatment-day imaging with contour uncertainty in adaptive proton radiotherapy

Yizhou Wu[1], Jie Ding[1], Justin Roper[1], Minglei Kang[2], Yuheng Li[1], Sibo Tian[1], David S. Yu[1], Xiaofeng Yang[3*], Chih-Wei Chang[1*]

[1]*Department of Radiation Oncology and Winship Cancer Institute, Emory University, Atlanta, GA 30308*

[2]*Department of Human Oncology, University of Wisconsin, Madison, WI 53792*

[3]*Department of Radiation and Cellular Oncology, University of Chicago, Chicago, IL 60637*

*Corresponding to: xfyang@uchicago.edu (XY) and chih-wei.chang@emory.edu (CC)

## Highlights

- A digital twin of prior-treatment deformations forecasts treatment-day HN anatomy
- Uncertainty is in the input, which library patient is followed, not in model weights
- Every replicate is a real deformation applied to the current patient's own anatomy
- The library fixes the anisotropic shape; the first QACT sets the scale of the band
- Six-direction contour uncertainty in millimeters for any delineated structure

## Abstract

Head-and-neck anatomy changes over a six-to-seven-week proton course, and the anatomy of a later week cannot be imaged when the plan is made. We present a digital-twin framework that forecasts a patient's treatment-day anatomy as an ensemble of predicted CTs with propagated contours and quantifies the uncertainty of the forecast contours. The twin is a library of previously treated patients with planning and weekly quality-assurance CTs (QACTs), made patient-specific by a two-step foundation-model deformable registration: a cross-patient field carries each library patient onto the current patient, and a longitudinal field, estimated in the current patient's frame, carries that patient's planning-to-QACT change onto the current patient's own planning CT. A library of 302 observations from 88 patients yields about 300 replicates per patient, each a deformation that occurred in a treated patient. The dispersion of the propagated contours, resolved by outward normal, is six-direction contour uncertainty in millimeters. This is uncertainty in the input to the forecast, which library patient the current patient follows, rather than in model parameters, and it is unchanged when the registration engine is exchanged. On ten patients with clinician contours on two QACTs, the library alone fixes the anisotropic shape of the uncertainty (4.5 to 6.2 mm); the first QACT narrows it by a factor of 3.2 to 3.6 without a contour being drawn; an approved contour improves the center but not the width. The estimate orders directions correctly but is not Gaussian-calibrated. A clinical target volume expansion is worked out as one application.

# 1. Introduction

Head-and-neck (HN) radiotherapy is delivered in 30 to 35 fractions over six to seven weeks, and the anatomy irradiated in the final week is not the anatomy that was planned (Barker et al., 2004). Gross tumor and parotid volumes fall throughout the course, with changes in external contour and organ position becoming pronounced after the third week (Barker et al., 2004). The parotid glands lose about 15% of their volume by week two and 31% by week four while shifting medially by 2 to 3 mm, and nodal targets and submandibular glands move by similar amounts (Bhide et al., 2010). Delivered on the original plan, these changes reduce target dose in most patients, raise the maximum dose to spinal cord and brainstem, and leave roughly 60% of parotid glands 4 Gy above their planned mean dose (Hansen et al., 2006). Proton therapy is more sensitive still, because the range of each pencil beam depends on the water-equivalent thickness along its path, so shrinkage that mildly perturbs a photon plan can move the distal edge of a proton field into an organ at risk or out of the target (Stützer et al., 2017). Replanning on the changed anatomy restores coverage and lowers organ-at-risk dose, and one well-timed replan captures most of the achievable benefit (Schwartz et al., 2013). The obstacle is its cost: re-imaging, re-contouring, re-optimization, quality assurance and physician approval occupy a multi-day care path even in a dedicated offline programme (Aristophanous et al., 2024). Meanwhile the patient is treated on a degraded plan or treatment is interrupted, and anatomical change is the most frequent trigger of adaptation in HN proton therapy. Nor is it clear in advance who will need adaptation: a systematic review found the predictors too heterogeneous to yield selection criteria (Brouwer et al., 2015).

Current practice manages this uncertainty by monitoring. Proton centers acquire weekly quality-assurance CTs (QACTs) and recalculate the plan on them, and these scans frequently drive the decision to replan (Evans et al., 2020; Stanforth et al., 2022); photon centers inspect the daily cone-beam CT (CBCT) or apply dose-based triggers derived from it (Aristophanous et al., 2024). Monitoring is reactive by construction: it detects change after it has occurred, the inspection is subjective, and each QACT adds imaging dose and scheduling load. Deep learning has been proposed to close this gap on the treatment day. Synthetic CT (sCT) networks correct the scatter, artifacts and Hounsfield-unit (HU) bias of CBCT so that dose can be calculated on the image of the day (Harms et al., 2019; Spadea et al., 2021), auto-segmentation networks delineate HN organs at risk and targets at near-expert accuracy (Nikolov et al., 2021), and CBCT-based evaluation frameworks combine the two to judge a plan within the treatment slot (Stanforth et al., 2022). Their limitations are documented. For proton Monte Carlo dose calculation, better image-quality metrics do not guarantee dosimetric agreement, because range depends on HU fidelity along the beam path, and generative networks occasionally introduce anatomically inconsistent features that shift range by millimeters (Thummerer et al., 2020). Auto-contours still need manual editing and sit at the edge of inter-observer variability (van Dijk et al., 2020). Most fundamentally, each tool returns a single image or contour, and where an uncertainty is attached it is obtained by perturbing the network, through dropout, ensembles or test-time augmentation, so it describes the model's confidence rather than the range of anatomies the patient may present (Gal and Ghahramani, 2016; Kendall and Gal, 2017). And all of it stays on the treatment-day critical path: synthesis, segmentation and optimization run after the image is acquired, with the patient on the table.

A digital twin (DT) offers a different route. A DT is a virtual replica of an individual that is updated with that individual's data and used to simulate and forecast (Katsoulakis et al., 2024; Venkatesh et al., 2024). In oncology, image-guided DTs have forecast patient-specific tumor response and optimized radiotherapy regimens under uncertainty (Asghar et al., 2025; Chaudhuri et al., 2023; Wu et al., 2022), and medical imaging has become the principal substrate for building them (Zhao et al., 2025). Our group has developed an anticipatory DT for adaptive radiotherapy in which a database of previously treated patients pre-generates the plans a new patient is likely to need, so that treatment-day adaptation reduces to selection; it preserved target coverage while shortening re-optimization in prostate stereotactic radiotherapy and showed dosimetric benefit in HN proton therapy (Chang et al., 2024, 2025, 2026a, 2026b). We then showed that a prior patient's planning-to-treatment change, transported onto the new patient through a cross-patient registration, predicts that patient's treatment-day anatomy more faithfully than the static planning CT (Wu et al., 2026). A DT is the appropriate instrument because the quantity to be predicted does not yet exist: the anatomy of week five cannot be imaged at planning, so no patient-specific model of it can be fitted, and the only evidence is how previously treated patients changed. Each library patient, deformably registered to the current patient, becomes a digital replicate of that patient undergoing one realized course of change; the library of replicates is the twin, and the variation across them is the uncertainty of the forecast. This is uncertainty in the input to the forecast, which library patient's change the current patient will follow, rather than in the form or the parameters of a fitted model (Chang and Dinh, 2019; Chang et al., 2020; Smith, 2013): model-parameter estimates shrink as models and training sets improve, whereas input uncertainty shrinks only with data about this patient. Those data arrive during treatment as weekly QACTs, and the twin is refined by retaining the replicates consistent with the observed anatomy. Because the replicates are generated before treatment, synthesis, contour propagation and plan preparation leave the treatment-day critical path.

Here we present a DT framework that forecasts the treatment-day anatomy of an HN patient as an ensemble of predicted CTs (pdCTs) with propagated organ-at-risk and target contours, and quantifies the uncertainty of the forecast contours. A two-step, foundation-model-based deformable registration first maps each library patient's planning CT onto the current patient, making the replicate patient-specific, and then extracts that patient's planning-to-QACT change in the current patient's frame; applying the change to the current patient's planning CT yields one pdCT and one contour set per library observation. The dispersion of the ensemble, parameterized on the consensus contour surface by outward normal, is reported as six-direction contour uncertainty in millimeters. The same construction applies to targets and to organs at risk, since both are delineated boundaries carried by the same deformation. The uncertainty is estimated first from the library alone and then refined when the patient's first QACT is available, and both estimates are validated against clinician contours on a later QACT. The contributions are (i) a formulation of anatomical-forecast uncertainty as a property of the prior-treatment data rather than of a model, with a DT estimator whose every sample is a deformation that occurred in a treated patient; (ii) the extension of cross-patient longitudinal transport (Wu et al., 2026) with patient-specific longitudinal refinement; (iii) a surface-normal parameterization converting the contour ensemble into direction-resolved millimeter uncertainty; (iv) a quantitative separation of what the library supplies, the anisotropic shape of the uncertainty, from what the patient's QACT supplies, its scale; and (v) a worked example of translation to a clinical target volume expansion for proton therapy, presented as one application of the methodology rather than its purpose. The framework is evaluated on a library of 302 planning-to-QACT observations from 88 HN patients and on ten patients with clinician-delineated structures on two QACTs each. Section 2 reviews related work on anatomical-change modeling and on uncertainty estimation in image synthesis, segmentation and registration; Section 3 presents the framework and experiments; Section 4 reports results; Sections 5 and 6 discuss and conclude.

## 2. Related work

### 2.1 Modeling and predicting inter-fraction anatomical change

Two families of methods anticipate how anatomy will change over a course. Image-conditioned methods use an image acquired at or early in treatment: a planning CT deformed to the daily CBCT yields a virtual CT whose replanning indicators match those of a rescan CT (Veiga et al., 2016), and synthetic-CT networks reconstruct a planning-quality image from the acquisition of the day (Spadea et al., 2021). These methods are accurate but conditional on an image that does not exist before treatment. Population models instead learn variability offline. Principal component analysis (PCA) of registration-derived deformation fields describes prostate, bladder, cervix and pancreatic organ motion with a few dominant eigenmodes, supports probabilistic dose-coverage evaluation, and can synthesise a planning library for a new patient (Budiarto et al., 2011; Söhn et al., 2005). For HN, DIR-based average and PCA models predict weekly anatomy and support anatomical robust optimization (Zhang et al., 2022), a Bayesian model combines a population prior with the patient's own early scans (Rørtveit et al., 2023), and recurrent, variational and diffusion networks generate probable future anatomies or tumor shrinkage trajectories (Pastor-Serrano et al., 2023; Smolders et al., 2024). These models share a design: they compress observed variation into parameters, a mean and eigenmodes or network weights, and then sample from the fitted distribution. The samples are statistically plausible but are not deformations that occurred, the models are typically fitted per organ or per region rather than for the whole anatomy with its contours, and, with the exception of Rørtveit et al. (2023), they do not update with the patient's own longitudinal data. Multi-atlas segmentation is the structural relative of what we propose: a library of registered exemplars is propagated onto a novel image and fused (Iglesias and Sabuncu, 2015). It transports labels of a fixed anatomy onto an image that exists; we transport deformations, changes over time, onto an anatomy that does not yet exist, and we treat the disagreement between library members as the quantity of interest rather than as noise to be fused away.

### 2.2 Uncertainty in deep-learning image synthesis

Uncertainty for sCT has been approached by making the synthesis network stochastic. Tanno et al. (2021) decomposed predictive uncertainty in neuroimage enhancement into an intrinsic and a parameter component and propagated it into downstream analysis. In radiotherapy, Monte Carlo dropout and ensemble uncertainty maps for synthetic CT have been shown to correlate with HU and proton-range error and have been proposed as quality-assurance tools and as inputs to robust optimization, and a systematic review finds dropout and ensembling to be the dominant estimators, applied mostly to auto-contouring, with little diversity in method and no standard for reporting (Wahid et al., 2024). In every case the source of variation is the network, and the question answered is how far the synthesized image may deviate from a truth that is available in principle on the day. The uncertainty we estimate has a different source and a different referent: it arises from which prior patient's change the current patient will follow, and it concerns an anatomy that cannot be imaged when the estimate is made.

### 2.3 Uncertainty in segmentation and in registration

Segmentation uncertainty is the most developed area. The taxonomy separates aleatoric uncertainty, irreducible variability in the data, from epistemic uncertainty about the model, and adds distributional uncertainty for test samples outside the training distribution (Hüllermeier and Waegeman, 2021; Kendall and Gal, 2017). Estimators randomise the network, through dropout, deep ensembles, test-time augmentation or a learned latent, as in the probabilistic U-Net (Gal and Ghahramani, 2016); calibration, out-of-distribution detection and the alignment of voxel-wise uncertainty with clinically relevant error have been studied systematically (Huang et al., 2024; Mehrtash et al., 2020). Registration uncertainty follows the same pattern. Bayesian non-rigid registration yields posterior distributions on the deformation that are non-Gaussian and multi-modal at ambiguous sites (Risholm et al., 2013), and probabilistic diffeomorphic networks learn a distribution over velocity fields (Dalca et al., 2019). In radiotherapy, DIR uncertainty is propagated into contour propagation and dose accumulation, learned directly by deep networks, integrated into online adaptive proton optimization, and governed by task-group and consensus recommendations (Nenoff et al., 2023). All of this quantifies how uncertain a registration or a segmentation is, given the images. None of it addresses how uncertain the future anatomy is, given only the planning image and a population.

### 2.4 The gap

Placed on the two axes of data-driven modeling, the prior knowledge available about a process and the data available about the case (Chang and Dinh, 2019), the estimators above occupy the high-knowledge corner: a model of the quantity is given, the data of the case are in hand, and what remains uncertain are the model's parameters. Anatomical forecasting before treatment occupies the opposite corner. No patient-specific model of the week-five anatomy can be fitted, because that anatomy has not occurred, and a library of previously treated patients has to stand in for it; the only prior knowledge in play is anatomical correspondence, supplied by the registration model, and everything about change comes from data. The classification of Smith (2013), adopted for proton dose calculation by Chang et al. (2020), separates three sources of systematic uncertainty: model form, the assumptions and implementation of the model; model parameters, its fitted or tabulated quantities; and input, the data fed to it. Dropout, ensembles and test-time augmentation estimate the second; substituting an architecture or a registration engine probes the first; the uncertainty that matters here is the third, which library observation the current patient will follow, and it is reducible only by data about this patient (Chang and Dinh, 2019). The DT framework of Section 3 estimates this input uncertainty directly: it keeps every library observation as a realized deformation rather than compressing it into parameters, transports it onto the current patient so that every sample is a deformation of this patient's own anatomy, attaches the propagated contours so that the uncertainty is available for any delineated structure, and refines the estimate as the patient's QACTs arrive; Section 4.5 tests its sensitivity to the model-form class by exchanging the registration engine.

## 3. Methodology

### 3.1 Overview

Figure 1 summarizes the framework. A prior-treatment library L holds, for each library patient j, a treatment planning CT $TPCT_j$ and the weekly quality-assurance CTs $QACT_{j,k}$ acquired during that patient's course. For a current patient c with planning CT $TPCT_c$ and planning contours $S_c$, every library observation (j,k) is transported onto patient c by two deformable registrations: a cross-patient field $DVF1_j$ makes library patient j patient-specific, and a longitudinal field $DVF2_{j,k}$, estimated in patient c's frame, carries the change that patient j underwent between planning and session k. Applying $DVF2_{j,k}$ to $TPCT_c$ and $S_c$ yields one replicate: a predicted CT $pdCT_{j,k}$ and its propagated contours $S_{j,k}$. The set of replicates is the digital twin $D_c$ of patient c, a patient-specific ensemble of anatomies that patient could present during treatment, each of which is a deformation that occurred in a treated patient. The dispersion of the propagated contours across $D_c$, resolved by outward normal direction on the consensus contour, is the contour uncertainty of the forecast. It is estimated from the library alone before treatment (library-prior uncertainty) and refined once the patient's first QACT is available by retaining the replicates whose images agree with it (QACT-refined uncertainty).

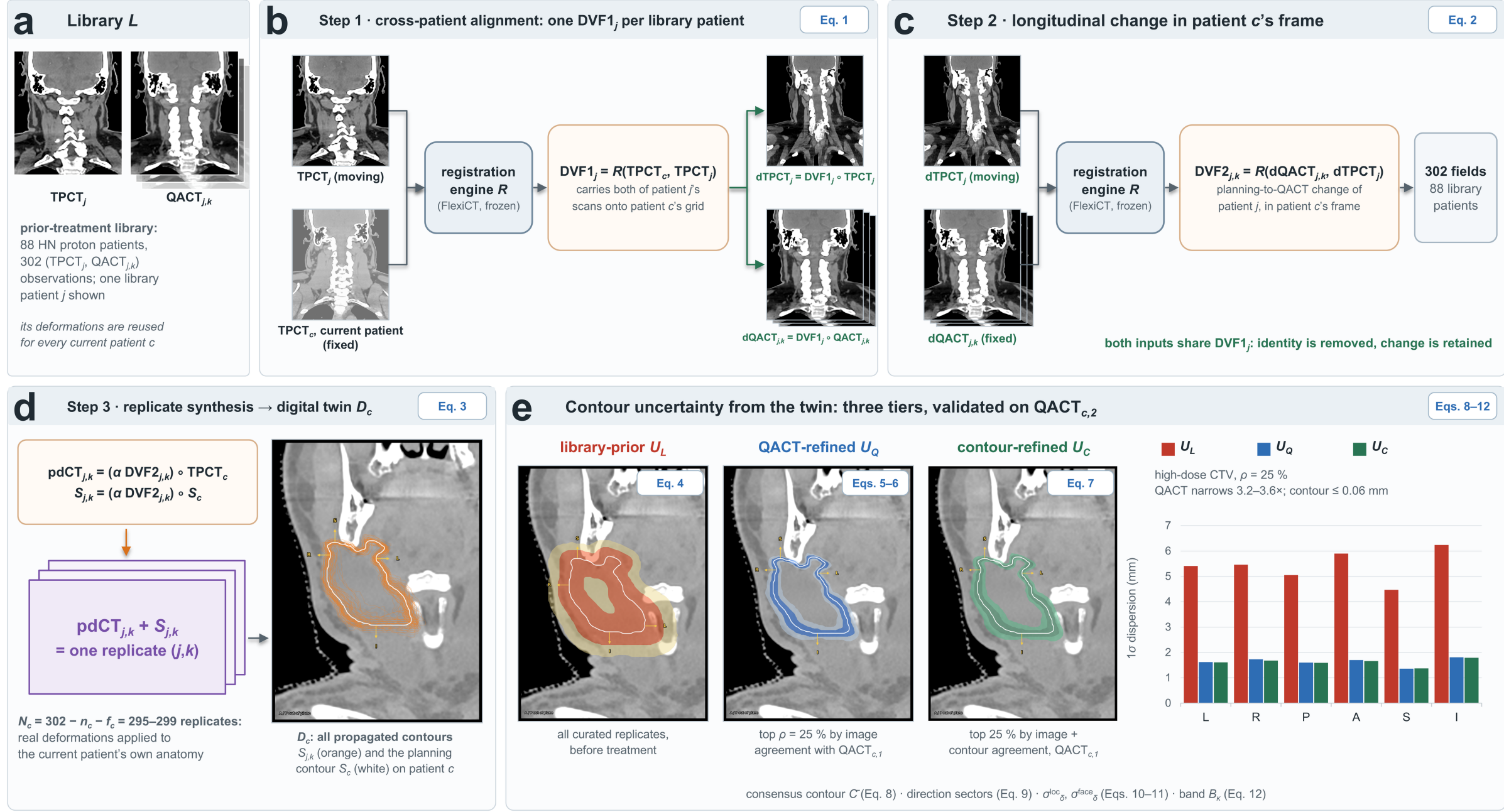

**Figure 1.** The digital-twin framework. (a) Prior-treatment library: each library patient j contributes a planning CT and weekly QACTs. (b) Step 1: the cross-patient field DVF1j carries both of patient j's scans onto the current patient's grid. (c) Step 2: the longitudinal field DVF2j,k is estimated between the deformed scans, in the current patient's frame. (d) Step 3: DVF2j,k is applied to the current patient's own planning CT and contours, giving one replicate (pdCTj,k, Sj,k); the set of replicates is the digital twin Dc. (e) Contour uncertainty from the twin: library-prior ($U_L$, all curated replicates) before treatment; QACT-refined ($U_Q$, replicates agreeing with $QACT_{c,1}$) and contour-refined ($U_C$, replicates agreeing with the approved contour on $QACT_{c,1}$) during treatment; all validated against clinician contours on $QACT_{c,2}$.

Table 1 lists the notation. Table 2 places the estimated quantity within the classification of systematic uncertainty into model-form, model-parameter and input components (Chang et al., 2020; Smith, 2013): the framework estimates input uncertainty, which library observation the current patient will follow, and Section 4.5 tests its sensitivity to model form by exchanging the registration engine. The cross-patient transport of Section 3.3 was introduced in our conference paper (Wu et al., 2026), which used the center of the ensemble to predict a single anatomy; the present work extends it with the longitudinal refinement of Section 3.4 and the contour-uncertainty quantification of Section 3.5, and treats the dispersion rather than the center as the deliverable.

**Table 1.** Notation used in the equations.

| Symbol | Meaning |
|---|---|
| *c*, *j* | current patient; library patient, $j = 1,\ldots,J$ |
| $TPCT_c$, $S_c$ | planning CT and planning contours of the current patient |
| $TPCT_j$, $QACT_{j,k}$ | planning CT and k-th weekly QACT of library patient j |
| $QACT_{c,1}$, $QACT_{c,2}$ | first and second QACT of the current patient; $S^{GT}_{c,k}$ the clinician contour on $QACT_{c,k}$ |
| $rTPCT_c$ | $TPCT_c$ rigidly aligned to the QACT frame (no-adaptation baseline) |

| Symbol | Meaning |
|---|---|
| R(F, M) | registration engine returning the field that aligns moving image M to fixed image F |
| $DVF1_j$, $DVF2_{j,k}$ | cross-patient and longitudinal deformation vector fields |
| $dTPCT_j$, $dQACT_{j,k}$ | library images warped into patient c's frame by $DVF1_j$ |
| $pdCT_{j,k}$, $S_{j,k}$ | predicted CT and propagated contours of replicate (j,k) |
| $\alpha$ | deformation scaling factor on DVF2 |
| $D_c$, K | digital twin (all replicates); retained set after curation and, where applied, refinement |
| $s_{img}$, $s_{ctr}$, $\rho$ | image-agreement score, contour-agreement score, retention fraction |
| $d_{j,k}(x)$, $\bar{d}(x)$, $\bar{C}$ | signed distance field of $S_{j,k}$ (negative inside), its mean over K, and the consensus contour $\{x : \bar{d}(x) = 0\}$ |
| n(x), $F_\delta$ | outward normal on $\bar{C}$; faces of $\bar{C}$ assigned to direction $\delta \in \{L,R,A,P,S,I\}$ |
| $\sigma^{loc}{}_\delta$, $\sigma^{face}{}_\delta$ | point-wise and rigid directional dispersion |
| $B_\kappa$ | directional band at multiplier $\kappa$ |

**Table 2.** Classes of systematic uncertainty (Smith, 2013; Chang et al., 2020) and where each appears in this work.

| Class | Source | Typical estimator | In this work |
|---|---|---|---|
| Model form | architecture, physics assumptions, registration engine | substitution of architecture or engine | Section 4.5: FlexiCT replaced by MIND |
| Model parameter | network weights, tabulated constants | Monte Carlo dropout, deep ensembles, test-time augmentation | the class addressed by the methods of Section 2 |
| Input | the data supplied to the model: which library observation the patient follows | the DT ensemble $D_c$ | Sections 3.4–3.5; the quantity reported in Section 4 |

## 3.2 Prior-treatment library and digital twin

The library comprises previously treated HN patients for whom a planning CT and one or more same-geometry QACTs exist. Each (patient, session) pair (j,k) is one longitudinal observation of how a treated anatomy changed between planning and week k. For a current patient c every library patient other than c is eligible, so J = 87 library patients contribute 302 - $n_c$ - $f_c$ observations, where $n_c$ is the number of patient c's own sessions (3 to 7) and $f_c$ the number of transport failures ($f_c$ = 0 for nine of the ten evaluation patients and 3 for one), giving 295 to 299 replicates per patient before curation. A single library patient cannot represent the current patient: two patients with the same planning anatomy lose weight at different rates and their tumors regress differently. The library as a whole can. Once each library patient is deformably registered to patient c (Section 3.3), that patient's course becomes a course that patient c's own anatomy could follow, and the collection of such courses is the digital twin $D_c$. Its purpose is twofold. Before treatment, the spread of $D_c$ bounds where each delineated boundary may lie on a future treatment day. During treatment, the patient's own QACT identifies the subset of $D_c$ consistent with the anatomy actually presented, so that the forecast for later sessions can be narrowed and the image sets, contours and, in the wider DT workflow, the plans prepared for that subset can be used without repeating synthesis, segmentation or optimization on the day (Chang et al., 2026a; Wu et al., 2026).

## 3.3 Two-step deformable image registration

Let R(F, M) denote the frozen, pretrained deformable registration engine, which returns the field that aligns moving image M to fixed image F, and let $\varphi \circ X$ denote warping an image or contour set X by field $\varphi$. Transport proceeds in three steps (Table 3, Eqs. 1–3).

*Step 1, cross-patient alignment.* One field per library patient carries both of that patient's scans onto the current patient's grid (Eq. 1). Because $TPCT_j$ and $QACT_{j,k}$ share the same field, inter-patient anatomical difference is removed from both while the difference between them, patient j's longitudinal change, is preserved. There are J such fields.

*Step 2, longitudinal change.* On the current patient's grid, a second registration estimates the change between the deformed planning and deformed QACT images of patient j (Eq. 2). Both inputs have passed through the same $DVF1_j$, so the estimated field encodes change rather than identity, and it is already expressed in patient c's frame. There is one $DVF2_{j,k}$ per library observation.

*Step 3, prediction.* Each transported change is applied to the current patient's own planning CT and planning contours (Eq. 3), the contours by nearest-neighbour interpolation of the same field. Mathematically the construction is a conjugation, $DVF2_{j,k} \approx DVF1_j \circ \psi_{j,k} \circ DVF1_j^{-1}$, where $\psi_{j,k}$ is patient j's longitudinal deformation in its own frame; the conjugation is what removes the library patient's identity and retains its change. To first order a displacement transforms under the conjugation as $u_c = J_{DVF1}^{-1} u_j$, so displacement vectors must be reoriented by the inverse Jacobian of the cross-patient field; Section 4.5 shows that omitting this term produces errors larger than the dispersion being estimated. A per-patient deformation scaling factor $\alpha$ multiplies DVF2 ($\alpha = 1.25$ for two patients by clinical direction and 1.0 otherwise); $\alpha = 0$ recovers the planning anatomy. Before scoring or evaluation, every $pdCT_{j,k}$ is rigidly aligned to the QACT frame on bone (HU clipped to [250, 1500], six degrees of freedom, multi-scale, normalized cross-correlation objective). The same alignment applied to $TPCT_c$ defines the no-adaptation baseline $rTPCT_c$, so that every reported gain is net of the setup correction that daily image guidance performs in any case.

## 3.4 Library-prior and QACT-refined uncertainty

Three tiers of the estimate are distinguished by the information they consume (Table 4). All three operate on the same digital twin; they differ only in which replicates are retained.

*Library curation.* A small number of library sessions produce implausible deformations on every recipient. They are identified once, offline, from the contoured QACTs of the retrospective cohort: a session is excluded if its replicates achieve a mean high-dose CTV Dice below 0.10 across recipients. The exclusion list is rebuilt under leave-one-patient-out cross-validation for every evaluated patient, so no information from the patient being evaluated enters its own curation. Thirteen sessions (4.4% of the library) are excluded, leaving 282 to 286 replicates per patient. Curation removes globally failing sessions; it does not select good ones. A contour-free alternative based on the influence of each replicate on the ensemble mean reproduces the list with an area under the ROC curve of 0.998 (Section 4.5), so the framework does not depend on contoured library data.

*Library-prior uncertainty ($U_L$).* No information from the current patient beyond $TPCT_c$ is used. The retained set is the curated twin, $K = D_c \setminus B$ (Eq. 4), with no ranking. This is the estimate available at planning.

*QACT-refined uncertainty ($U_Q$).* When the patient's first QACT is acquired, each replicate's predicted image is compared with it by the image-agreement score $s_{img}$, the equal-weight mean of normalized cross-correlation, structural similarity and inverse mean absolute error over the body mask (Eq. 5), after the rigid bone alignment of Section 3.3. Replicates are ranked and the top fraction $\rho$ is retained (Eq. 6). No contour is drawn on the QACT; the operations performed on it, rigid setup and intensity similarity, are those daily image guidance already performs. We use $\rho = 0.25$ as the operating point and evaluate that choice in Section 4.4 by sweeping $\rho$ and by validating the refined estimate against the clinician contour on the patient's *second* QACT, so that refinement uses $QACT_{c,1}$ and validation uses $QACT_{c,2}$. Restricting $s_{img}$ to a region of interest around the planning target, or replacing it with normalized cross-correlation alone, moves the estimate by at most 0.05 mm in any direction (Section 4.5), so the conclusions do not depend on the region over which agreement is scored.

*Contour-refined uncertainty ($U_C$).* If, in addition, a physician-approved contour exists on $QACT_{c,1}$, the score gains two contour terms, the Dice coefficient and a clipped average symmetric surface distance between each replicate's propagated contour and the approved one (Eq. 7), and the top $\rho$ by this five-term score is retained. The contour terms account for 73% of the score variance, so $U_C$ is in effect selection by agreement with the approved contour. $U_C$ is the most informed tier and the upper bound on what same-patient information can contribute; it is reported to separate what the QACT image supplies from what an approved contour on it would add.

### 3.5 Contour-uncertainty quantification

*Aggregation.* Let K be the retained set of size N and $d_{j,k}(x)$ the signed distance field of the propagated contour $S_{j,k}$, negative inside. The ensemble is summarized by the mean signed distance field $\bar{d}(x)$ and its zero level set, the consensus contour $\bar{C}$ (Eq. 8). Averaging signed distance fields of mutually displaced contours contracts the zero level set, so $\bar{C}$ is systematically smaller than its members; the amount depends on the spread of the retained set and is reported in Section 4 wherever a result depends on the center. The dispersion is measured relative to the members and is unaffected.

*Direction sectors.* The natural parameterization, casting rays from the centroid of $\bar{C}$ along the six cardinal axes, fails for HN structures: elective nodal volumes form a collar around the airway whose centroid lies in air, and 20% of structure-by-direction combinations (39.6% at the lowest dose level) have no intersection (Supplementary Fig. S1). We therefore parameterize on the surface itself. $\bar{C}$ is triangulated, the outward normal $n(x) = \nabla \bar{d} / \| \nabla \bar{d} \|$ is evaluated on every face, and each face is assigned to the direction $\delta \in \{L, R, A, P, S, I\}$ whose unit vector $e_\delta$ maximizes $n(x) \cdot e_\delta$ (Eq. 9). This is defined on every face by construction and solved all 230 structure-by-level and 1,380 structure-by-level-by-direction combinations in the cohort.

*Directional dispersion.* The displacement of replicate (j,k) at a point x of $\bar{C}$ is $u_{j,k}(x) = -d_{j,k}(x)$, positive where the replicate lies outside $\bar{C}$. Two statistics are retained per direction sector $F_\delta$, both area-weighted (Eqs. 10–11): $\sigma^{loc}{}_{\delta}$, the point-wise standard deviation of $u_{j,k}(x)$ across replicates, which includes shape change within the sector; and $\sigma^{face}{}_{\delta}$, the standard deviation across replicates of the sector-mean displacement, which treats the sector as a rigid translation. The band uses $\sigma^{loc}{}_{\delta}$. Their ratio $r = \sigma^2{}_{face} / \sigma^2{}_{loc}$ bounds, by Jensen's inequality, the fraction of contour-uncertainty variance that any per-axis constant field can represent; it is measured in Section 4.6. The mean directional displacement across the retained set is at most 0.009 mm in magnitude, so the ensemble is unbiased with respect to its own members and the uncertainty is dispersion, not offset.

*Directional band.* For display and for the translational example, the band at multiplier $\kappa$ is the set of voxels within $\kappa\, \sigma^{loc}{}_{\delta(x)}$ of the center (Eq. 12), where $\delta(x)$ is the direction bucket of the nearest surface point. This is an anisotropic expansion in physical distance; a uniform expansion is the special case in which all six $\sigma^{loc}{}_{\delta}$ are replaced by one constant. The correspondence between $\kappa$ and geometric coverage is established empirically in Section 4.4 rather than assumed Gaussian. All reported values use the mean-based aggregate.

**Table 3.** Equations of the framework and the panel of Figure 1 in which each step appears.

| No. | Equation | Description | Framework step (Figure 1) |
|---|---|---|---|
| (1) | $\mathrm{DVF1_j} = R(\mathrm{TPCT_c}, \mathrm{TPCT_j});\ \mathrm{dTPCT_j} = \mathrm{DVF1_j} \circ \mathrm{TPCT_j};\ \mathrm{dQACT_{j,k}} = \mathrm{DVF1_j} \circ \mathrm{QACT_{j,k}}$ | cross-patient alignment (one field per library patient) | Fig. 1(b), step 1: cross-patient alignment |
| (2) | $\mathrm{DVF2_{j,k}} = R\left(\mathrm{dQACT_{j,k}}, \mathrm{dTPCT_j}\right)$ | longitudinal change in the current patient's frame (one field per observation) | Fig. 1(c), step 2: longitudinal field |
| (3) | $\mathrm{pdCT_{j,k}} = (\alpha\, \mathrm{DVF2_{j,k}}) \circ \mathrm{TPCT_c};\ S_{\mathrm{j,k}} = (\alpha\, \mathrm{DVF2_{j,k}}) \circ S_{\mathrm{c}}$ | replicate: predicted CT and propagated contours | Fig. 1(d), step 3: replicate generation → twin Dc |
| (4) | $K_{\mathrm{L}} = D_{\mathrm{c}} \setminus B$ | library-prior retained set (B: curated sessions) | Fig. 1(e), library-prior tier $U_L$ |
| (5) | $s_{\mathrm{img}}(\mathrm{j,k}) = \frac{1}{3}\left[\frac{\mathrm{NCC}+1}{2} + \frac{\mathrm{SSIM}+1}{2} + \frac{1}{1+\mathrm{MAE}}\right]$ over the body mask of $\mathrm{QACT_{c,1}}$ | image-agreement score | Fig. 1(e), QACT-refined tier $U_Q$: scoring |
| (6) | $K_{\mathrm{Q}} = \{(\mathrm{j,k}) \in K_{\mathrm{L}}: \mathrm{rank}\, s_{\mathrm{img}}(\mathrm{j,k}) \le \rho\lvert K_{\mathrm{L}}\rvert\}$ | QACT-refined retained set | Fig. 1(e), QACT-refined tier $U_Q$: retention |
| (7) | $s_{\mathrm{ctr}}(\mathrm{j,k}) = \frac{1}{5}\left[\text{three terms of (5)} + \mathrm{Dice}\left(S_{\mathrm{j,k}}, S_{\mathrm{c,1}}^{\mathrm{GT}}\right) + \mathrm{clip}\left(1 - \frac{\mathrm{ASSD}\left(S_{\mathrm{j,k}}, S_{\mathrm{c,1}}^{\mathrm{GT}}\right)}{\delta_0}, 0,1\right)\right]$ | contour-agreement score ($\delta_0$ = 10 mm; the ASSD term is clipped to [0, 1]) | Fig. 1(e), contour-refined tier $U_C$: scoring |
| (8) | $\bar{d}(x) = \frac{1}{N}\sum_{(\mathrm{j,k})\in K} d_{\mathrm{j,k}}(x);\ \bar{C} = \{x: \bar{d}(x) = 0\}$ | mean signed distance field; consensus contour | Fig. 1(e), uncertainty: consensus contour |
| (9) | $\delta(x) = \underset{\delta\in\{\mathrm{L,R,A,P,S,I}\}}{\arg\max}\ n(x)\cdot e_{\delta}$ | direction assignment on $\bar{C}$ | Fig. 1(e), uncertainty: direction sectors |

| No. | Equation | Description | Framework step (Figure 1) |
|---|---|---|---|
| (10) | $(\sigma_\delta^{\mathrm{loc}})^2 = \frac{\sum_{x \in F_\delta} a(x)\, \mathrm{Var}_{(\mathrm{j,k})}\left[u_{\mathrm{j,k}}(x)\right]}{\sum_{x \in F_\delta} a(x)}$ | point-wise directional dispersion (a: face area) | Fig. 1(e), uncertainty: dispersion |
| (11) | $(\sigma_\delta^{\mathrm{face}})^2 = \mathrm{Var}_{(\mathrm{j,k})}\left[\bar{u}_{\mathrm{j,k},\delta}\right],\quad \bar{u}_{\mathrm{j,k},\delta} = \frac{\sum_{x \in F_\delta} a(x)\, u_{\mathrm{j,k}}(x)}{\sum_{x \in F_\delta} a(x)}$ | rigid directional dispersion | Fig. 1(e), uncertainty: dispersion |
| (12) | $B_\kappa = \left\{x: \left\lvert\bar{d}(x)\right\rvert \le \kappa\, \sigma_\delta^{\mathrm{loc}}(x)\right\}$ | directional band at multiplier $\kappa$ | Fig. 1(e), uncertainty: directional band |

**Table 4.** The three uncertainty tiers.

| Tier | When available | Information beyond $\mathrm{TPCT}_c$ | Retained replicates | Role |
|---|---|---|---|---|
| Library-prior $U_L$ | at planning | prior-treatment library; curation list from other patients | all curated (282–286), unranked | pre-treatment estimate; shape of the anisotropy; comparator for population constants |
| QACT-refined $U_Q$ | after $\mathrm{QACT}_{c,1}$ | $U_L$ + rigid setup and image agreement with $\mathrm{QACT}_{c,1}$ | top $\rho$ = 25% by $s_{img}$ | primary reported estimate; validated on $\mathrm{QACT}_{c,2}$ |
| Contour-refined $U_C$ | after physician approval of a contour on $\mathrm{QACT}_{c,1}$ | $U_Q$ + Dice and ASSD to the approved contour | top 25% by $s_{ctr}$ | upper bound on same-patient information; separates image from contour contribution |

### 3.6 Experiments

*Data.* A retrospective cohort of 88 HN patients treated with proton therapy at a single institution, each with a planning CT and same-geometry QACTs acquired during treatment, forms the library of 302 observations. Table 5 summarizes the cohort and implementation. Contour analyses are carried out on an evaluation subset of ten patients with clinician-delineated clinical target volumes (CTVs) at three dose levels on two QACTs each; the high-dose CTV is the primary structure. Organ-at-risk contours were available for all ten evaluation patients on both QACTs, covering the eight structures carried in the aligned ground truth, including the spinal cord, larynx, mandible, brainstem, both parotid glands, the oral cavity and the esophagus, and the right parotid and brainstem are used to illustrate the construction. Voxel spacing is 0.977 × 0.977 × 1.0 mm and all analyses are at full resolution. One evaluation structure is flagged: in one patient the high-dose CTV measures 50.4 cc at planning against 135.3 cc on the QACT, a ratio of 2.68 where the other nine lie between 0.54 and 1.42; Section 4.4 shows this is a target-definition ceiling rather than a forecasting failure, and the unit is retained.

*Implementation.* Both registrations use the deformable interface of the FlexiCT CT foundation model (Li et al., 2026) without patient-specific training (500 iterations, learning rate 6.0, smoothness weight 1.0, feature-map downsampling 2, iterative smoothing kernel 7 applied twice, full-resolution upsampling of the final field). Rigid alignment to the QACT frame uses a six-degree-of-freedom Euler transform centered on the QACT bone centroid, normalized cross-correlation with 25% sampling, regular-step gradient descent (400 iterations) and shrink factors 4/2/1 with smoothing sigmas 2/1/0 mm. The MIND descriptor (Heinrich et al., 2012) is substituted for the FlexiCT features in the engine-substitution test. Replicates are warps with propagated contours; no dose is calculated in this work, and how many replicates are dose-optimized in the wider DT workflow is a resource decision outside its scope. Offline generation of the twin takes 32 to 79 min per patient on one GPU; scoring against a QACT takes 1 to 2 min.

*Design of the refinement test.* Refinement uses $QACT_{c,1}$ and validation uses the clinician contours $S^{GT}{}_{c,2}$ on $QACT_{c,2}$, giving ten validation units for $U_Q$ and $U_C$. $U_L$ uses no patient data and is validated on both QACTs (20 units).

*Evaluation criteria.* Contour accuracy of the consensus contour is the Dice coefficient against $S^{GT}$; the no-adaptation reference is the planning contour on $rTPCT_c$. Rank agreement between the estimated dispersion and the realized error is the within-unit Spearman correlation between the six $\sigma^{loc}{}_{\delta}$ and the six root-mean-square mean-centered signed residuals of $\bar{C}$ from $S^{GT}$. Calibration is the area-weighted fraction of the $S^{GT}$ surface lying within $B_{\kappa}$ for $\kappa = 1, 2, 3$, compared with the Gaussian expectation; the distribution of the signed residual is reported by direction. Two coverage criteria appear and are not interchangeable: geometric surface coverage, as just defined, and volumetric coverage, the fraction of ground-truth volume inside an expanded structure, used only in the translational example; neither is a dosimetric criterion. Paired comparisons use the Wilcoxon signed-rank test with the count of units in which one condition exceeds the other; where two QACTs of one patient enter, p-values are also reported clustered by patient, and where the two disagree both are given.

*Robustness experiments. Six are run and reported in Section 4.5: omission of the inverse-Jacobian reorientation term; computation at half resolution, with a sphere-jitter model of quantization noise; the region and form of the image-agreement score (body mask against CTV neighbourhood; three terms against NCC alone); library size, by subsampling the curated twin to n library patients with 20 random draws per n; substitution of the registration feature extractor on one QACT across the ten patients; and replacement of the contour-based curation by a contour-free influence statistic. The retention fraction ρ is swept from 25% to 100% in Section 4.4.*

**Table 5.** Cohort and implementation summary.

| Item | Value |
| --- | --- |
| Library | 88 HN proton patients; 302 (patient, session) observations |
| Replicates per current patient | 295–299 before curation; 282–286 after |
| Curation | 13 sessions (4.4%) excluded; leave-one-patient-out; threshold mean high-dose CTV Dice < 0.10 |
| Evaluation cohort | 10 patients × 2 QACTs; CTV at three dose levels (20 high-dose, 20 mid-dose, 6 low-dose structure-QACT units) |
| Refinement / validation | $QACT_{c,1}$ / $S^{GT}{}_{c,2}$ (10 units for $U_Q$, $U_C$); $U_L$ on both QACTs (20 units) |
| Voxel spacing | 0.977 × 0.977 × 1.0 mm, full resolution |

| Item | Value |
|---|---|
| Registration engine | FlexiCT deformable interface (primary); MIND (substitution test) |
| Deformation scaling $\alpha$ | 1.25 for two patients (clinical direction), 1.0 otherwise |
| Retention $\rho$ | 0.25 (operating point); 0.25–1.0 swept |
| Compute | twin generation 32–79 min/patient (1 GPU); QACT scoring 1–2 min/patient |

### 3.7 Translational example: clinical target volume expansion

As one application of the contour uncertainty, the directional band around the CTV is compared with the uniform expansion used in HN proton practice, where 3 mm is the floor permitted under daily image guidance (van Herk et al., 2000). For each evaluation unit the planning high-dose CTV on $rTPCT_c$ is expanded uniformly by r mm, and the directional band $B_\kappa$ of Eq. 12 is swept over $\kappa$ with center and dispersion taken from each tier in turn. Because the same r produces very different volumes in different patients, the two are compared through the volume they produce: volumetric coverage of $S^{GT}$ is plotted against the ratio of expanded volume to ground-truth volume, interpolated on each unit's own curve before aggregation, and the arms are compared pairwise within units at matched volume and at matched coverage. A two-by-two factorial over center (planning CTV or consensus contour) and shape (uniform or directional) separates the contribution of each. The estimate covers inter-fraction anatomical deformation only; setup, delineation and dose-calculation uncertainties are not included, and the coverage criterion is geometric rather than the population-dosimetric criterion of margin recipes, so no recommendation on the size of clinical margins is made.

## 4. Results

### 4.1 Fidelity of the predicted CT

The images on which every later result rests are characterized first (Figure 2). Coronal slices of the QACT, the unadapted planning CT ($rTPCT_c$) and a replicate's pdCT are shown for two units at two levels fixed by rule (40th and 80th percentile of the high-dose CTV extent), with the body-voxel HU distribution of each slice. The predicted image reads as an ordinary HN CT whose body outline, airway and bone follow the anatomy of the day, while $rTPCT_c$ retains the planning anatomy. The Wasserstein-1 distance between each slice's HU histogram and that of the QACT is 10.1 to 16.5 HU for $rTPCT_c$ and 13.6 to 15.8 HU for pdCT, the same order and small against the soft-tissue HU range. Over 3,470 coronal slices in 20 units, the median body HU of pdCT deviates from the QACT by +1.63 HU (against +1.33 HU for $rTPCT_c$), the 95th-percentile HU by −11.48 HU as a constant offset, and the local noise is 19% lower (per-unit median ratio 0.81; −0.97 HU) in all 20 units; the noise deficit is traced to the two trilinear resamplings of the generation pipeline against one for the acquired images, and composing them into a single mapping restores the noise to acquired levels (ratio 0.97–1.05 in a three-unit check). The magnitude of predicted change matches the magnitude of actual change, a median slice-wise ratio of 0.98 over 4,664 slices, and the signed change fields correlate voxel-wise at a median r of 0.49 (0.24–0.77) across units. The predicted CT is therefore an image of the same class a clinic already works from, moved by the right amount with about half of the movement placed correctly; geometric accuracy is the subject of the next subsections. Image-agreement statistics of the retained sets against the QACT are given in Supplementary Table S1. Agreement rises monotonically with selectivity, from the whole curated pool through the top quarter to the single highest-ranked replicate, so the image-agreement score orders library members correctly across QACTs. The retained quarter nonetheless does not exceed the unadapted planning CT on these whole-body metrics, and the highest-ranked replicate exceeds it only marginally; such a metric is dominated by the anatomy the two images share, and the refinement is not directed at it. What the refinement is for is the contour and dispersion results of the following subsections.

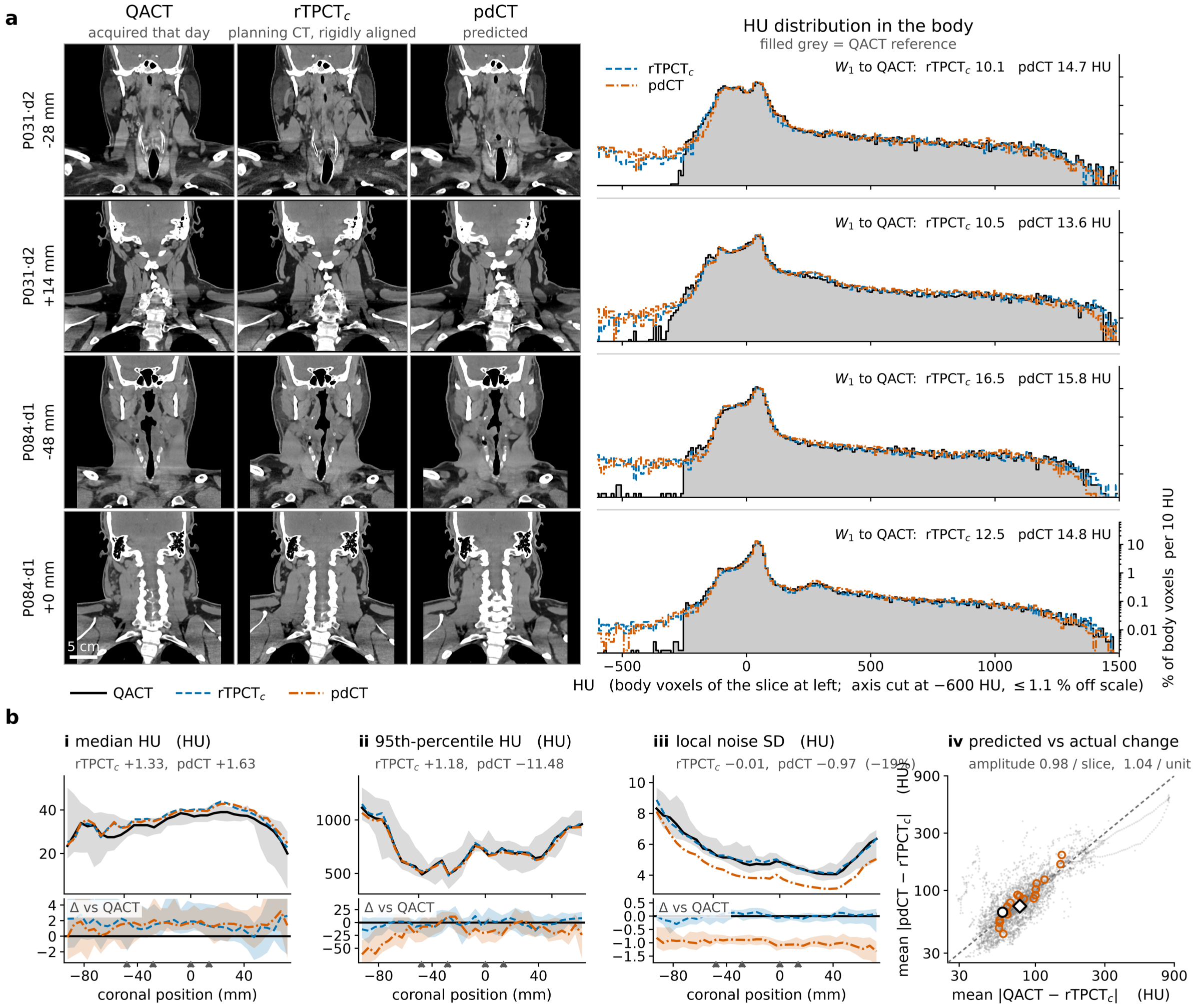

**Figure 2**. Fidelity of the predicted CT. (a) Coronal QACT, $rTPCT_c$ and pdCT for two patient–QACT units at two levels, soft-tissue window W/L 400/50 HU, with each slice's body-voxel HU distribution and its Wasserstein-1 distance to the QACT histogram. (b) Per-slice median HU, 95th-percentile HU, local noise and predicted-versus-actual change along the coronal axis, 20 units.

## 4.2 Contour propagation: the consensus contour

Before the dispersion can be read, the center of the twin must be located relative to the planning contour (Table 6). The uncurated twin is unusable: a few implausible replicates dominate the mean, and the consensus volume collapses to 51% below the ground truth. After curation, the library-prior consensus contour is statistically indistinguishable from the unadapted planning contour (Dice 0.7391 against 0.7436; 9 of 20 units; $p = 0.55$). The library therefore does not improve the point forecast; what it supplies before treatment is the dispersion around a center no better than the planning CT. QACT refinement improves the center by +0.0152 Dice (8 of 10 units, $p < 0.05$), roughly half the way to the contour-refined tier, which improves it by +0.0267 in every unit; the difference, +0.0115 Dice, is what a physician-approved contour adds to the center. Averaging signed distance fields contracts the consensus contour (Section 3.5) by an amount that depends on the spread of the retained set: at zero margin the library-prior consensus contour holds 0.80 of the ground-truth volume and the QACT-refined contour 0.89. Every quantity built on the center inherits this contraction; the dispersion does not.

**Table 6.** Dice of the consensus contour against the clinician contour, by tier, high-dose CTV unless stated. Reference: planning contour on $rTPCT_c$, 0.7436 over the twenty units on which the library-prior tier is validated and 0.7170 over the ten forward units of the refined tiers.

| Tier | Information beyond $TPCT_c$ | CTV level | Consensus Dice | $rTPCT_c$ | Gain | Units won |
|---|---|---|---|---|---|---|
| Uncurated twin | none | high | 0.5745 | 0.7436 | −0.1691 | 2/20 |
| Library-prior $U_L$ | none from this patient | high | 0.7391 | 0.7436 | −0.0045 | 9/20 ($p$ = 0.55) |
| QACT-refined $U_Q$, $\rho = 25\%$ | $QACT_{c,1}$ image | high | 0.7322 | 0.7170 | +0.0152 | 8/10 ($p <$ 0.05) |
| Contour-refined $U_C$, $\rho = 25\%$ | approved contour on $QACT_{c,1}$ | high | 0.7437 | 0.7170 | +0.0267 | 10/10 ($p <$ 0.01) |
| Contour-refined $U_C$, $\rho = 25\%$ | approved contour on $QACT_{c,1}$ | mid | 0.8063 | 0.7805 | +0.0258 | 10/10 ($p <$ 0.01) |
| Contour-refined $U_C$, $\rho = 90\%$ | approved contour on $QACT_{c,1}$ | high | 0.7276 | 0.7170 | +0.0106 | 7/10 ($p$ = 0.232) |

### 4.3 Contour uncertainty: library-prior versus QACT-refined

Direction-specific dispersion of the high-dose CTV is given in Table 7 for the three tiers, per-unit distributions in Figure 3, and the band drawn on the anatomy in Figure 4. Three properties are separated. First, the *shape* of the uncertainty is a property of the library and of this patient's anatomy. The library-prior band is already anisotropic, tightest superiorly (4.47 mm) and widest inferiorly (6.24 mm), and the six directions keep the same rank order under all three tiers (S < P < L < R < A < I); under QACT refinement two adjacent pairs, P/L and R/A, are separated by less than 0.01 mm; the inferior-to-superior ratio is 1.40 (1.33 and 1.27 under the refined tiers). Across dose levels the refined band is between 3% narrower and 16% wider at the low-dose than at the high-dose level under QACT refinement, and 3 to 27% wider under contour refinement, whereas the library-prior band is 8 to 28% narrower there (Supplementary Table S2). Second, the patient's QACT sets the *scale*. Relative to $U_L$, the QACT-refined band is narrower by a factor of 3.2 to 3.6 in every direction (L 3.38, R 3.30, P 3.17, A 3.56, S 3.33, I 3.50), obtained without any contour on the QACT; this is the quantity that measures what the first QACT contributes to the forecast. Third, an approved contour adds nothing to the width: the contour-refined band lies within 0.06 mm of the QACT-refined band in every direction, while moving the center by +0.0115 Dice (Table 6). The width is fixed by the image; the contour improves only the center.

Across patients the dispersion is not a constant (Table 8). Within the high-dose CTV under QACT refinement the ratio of the largest to the smallest per-unit $\sigma^{loc}{}_{\delta}$ is at least 1.79 in every direction and reaches 2.55 on the left, where one unit requires 2.14 mm and another 0.84 mm; under $U_C$ the ratios are 1.64 to 2.11 and under $U_L$ 3.97 to 5.03. The spread is largest before refinement and is reduced, not created, by the QACT. It is largest at the high-dose level, which is small, adjacent to the gross tumor and in the region of greatest change, and smallest at the low-dose level (max/min 1.05 to 1.49 under $U_C$,

n = 3). Together with a left–right asymmetry that reverses sign between patients, this is the direct evidence that no single population constant describes the cohort.

**Table 7.** Direction-specific 1-$\sigma$ dispersion $\sigma^{loc}{}_{\delta}$ (mm) of the high-dose CTV, mean over units, $\rho$ = 25% for the refined tiers. L/R/P/A/S/I: left, right, posterior, anterior, superior, inferior.

| Tier | L | R | P | A | S | I |
|---|---|---|---|---|---|---|
| Library-prior $U_L$ | 5.41 | 5.46 | 5.05 | 5.90 | 4.47 | 6.24 |
| QACT-refined $U_Q$ | 1.60 | 1.66 | 1.60 | 1.66 | 1.34 | 1.78 |
| Contour-refined $U_C$ | 1.58 | 1.61 | 1.57 | 1.65 | 1.36 | 1.72 |
| Narrowing $U_L$ / $U_Q$ | 3.38 | 3.30 | 3.17 | 3.56 | 3.33 | 3.50 |

**Table 8.** Inter-patient spread of $\sigma^{loc}{}_{\delta}$ (mm), high-dose CTV: min / median / max and max/min ratio by tier.

| Direction | $U_L$ | max/min | $U_Q$ | max/min | $U_C$ | max/min |
|---|---|---|---|---|---|---|
| L | 1.97 / 5.28 / 9.90 | 5.03 | 0.84 / 1.55 / 2.14 | 2.55 | 1.19 / 1.49 / 2.30 | 1.93 |
| R | 2.37 / 5.21 / 9.80 | 4.13 | 0.87 / 1.78 / 2.10 | 2.42 | 1.09 / 1.60 / 2.31 | 2.11 |
| P | 2.11 / 4.81 / 9.35 | 4.43 | 0.87 / 1.68 / 2.19 | 2.51 | 1.09 / 1.61 / 2.06 | 1.89 |
| A | 2.52 / 5.82 / 10.61 | 4.22 | 1.10 / 1.63 / 2.23 | 2.02 | 1.21 / 1.68 / 1.99 | 1.64 |
| S | 1.97 / 4.02 / 9.57 | 4.87 | 0.99 / 1.30 / 1.77 | 1.79 | 1.05 / 1.34 / 1.74 | 1.66 |
| I | 2.64 / 6.15 / 10.49 | 3.97 | 0.99 / 1.77 / 2.45 | 2.47 | 1.27 / 1.70 / 2.10 | 1.65 |

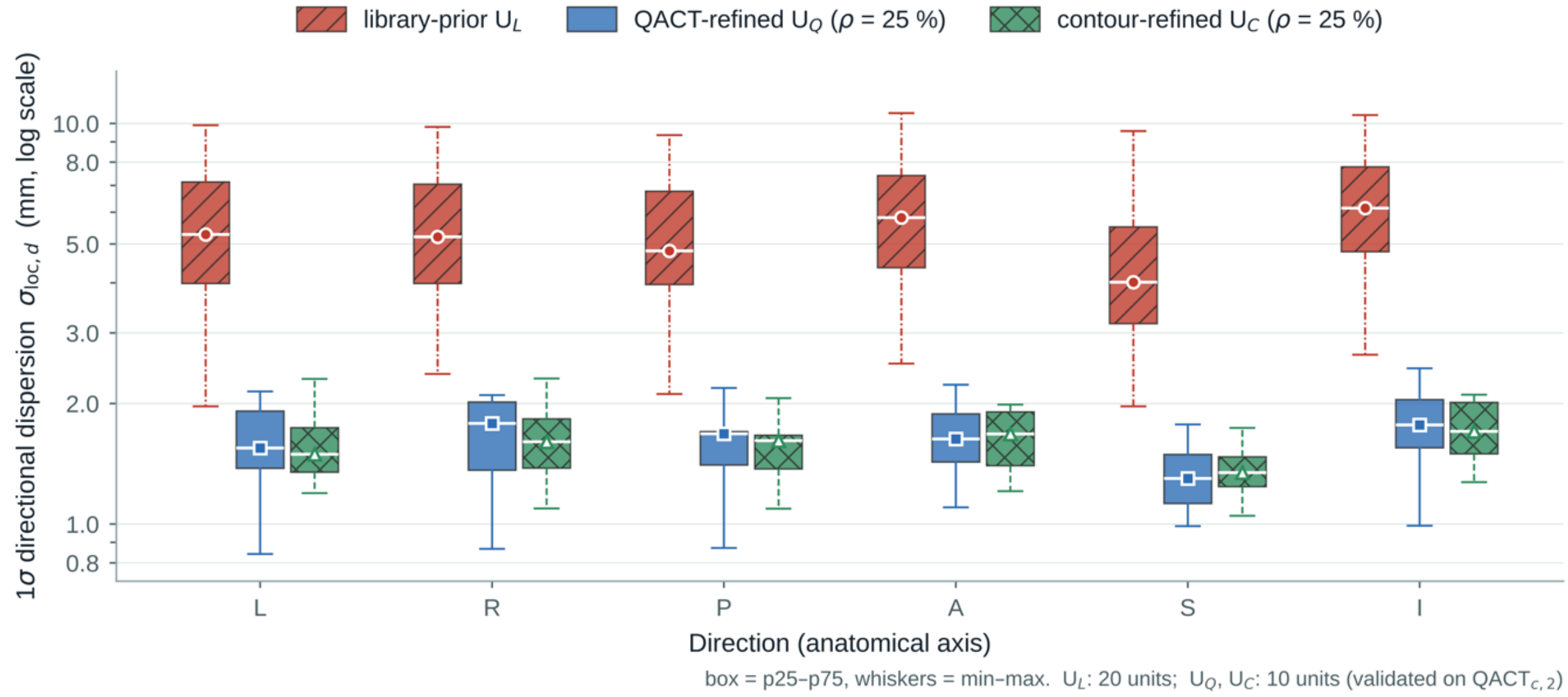

**Figure 3.** Direction-specific 1-$\sigma$ dispersion per unit (log scale) under the three tiers, refined tiers at $\rho$ = 25%. Boxes: inter-quartile range and median; whiskers: full range. Superior is the tightest and inferior the widest direction in every tier.

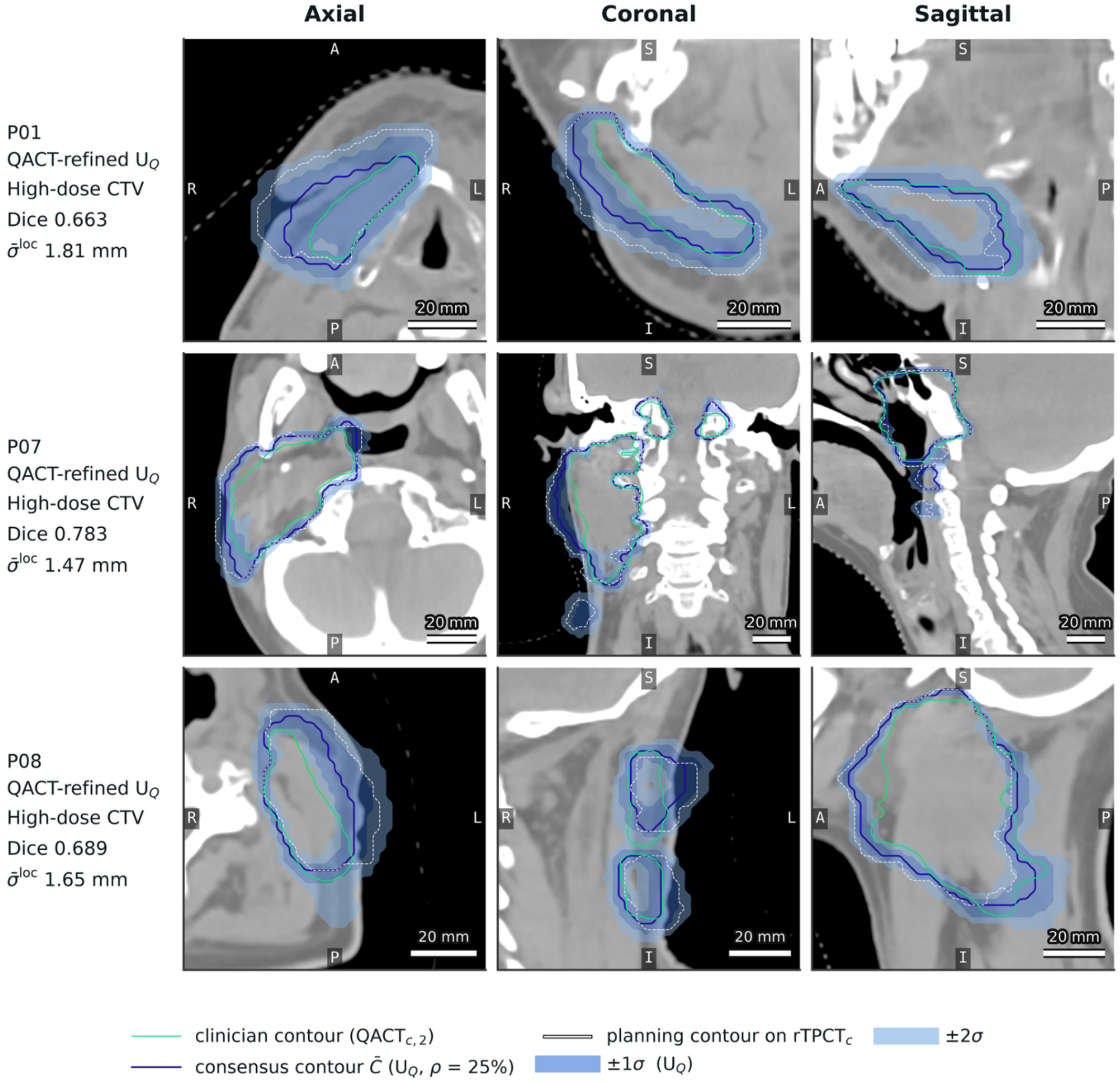

**Figure 4.** The directional band on the treatment-day anatomy, QACT-refined tier: the high-dose CTV of three patients, each in three orthogonal planes. Green, clinician contour; dark blue, consensus contour; dashed, planning contour on $rTPCT_c$; shaded, $\pm 1\ \sigma$ and $\pm 2\ \sigma$ directional bands. Each row gives that patient's own consensus Dice and the mean of $\sigma^{loc}$ over the six directions, 1.47 to 1.81 mm here. The three were chosen so that the consensus contour is at least one voxel closer to the clinician contour than the planning contour in every panel, which is what the figure is meant to show; the spread of $\bar{\sigma}^{loc}$ across the whole cohort is in Table 8. The same construction drawn on eight organs at risk is in Figure 5(d).

*Organs at risk.* These results use the design of the target results: refinement on $QACT_{c,1}$, validation against the clinician contour on $QACT_{c,2}$, ten patients, $\rho$ = 25% (Figure 5). The center improves significantly for five of the eight structures, including the mandible by 0.087 Dice (10 of 10 units, $p < 0.01$), the right parotid by 0.087, the brainstem by 0.086, the left parotid by 0.069 and the oral cavity by 0.043 (9 of 10, $p < 0.01$), while the larynx and the spinal cord gain 0.032 and 0.026 without reaching significance and the esophagus gains nothing (Table 9). The two thin tubular structures behave differently from the rest: under the library prior alone the esophagus loses 0.215 Dice and its average symmetric surface distance rises from 2.26 to 7.78 mm (2 of 10, $p < 0.05$), and the spinal cord also falls below the unadapted contour, which is the contraction of the zero level set that averaging signed distance fields produces when the thickness of a structure is comparable with the dispersion. Width behaves as it does for the target (Table 10): the library-prior band is 5.01 to 7.62 mm on the right parotid and 7.96 to 9.46 mm on the brainstem, the treatment-day image narrows it by a factor of 3.69 to 5.41 and 8.76 to 10.41 respectively, more than the 3.17 to 3.56 measured on the high-dose CTV, and the approved contour narrows it further by at most 0.10 mm in any direction on either structure. The anisotropy is not the target's: on the right parotid the refined band is widest laterally (R, 1.86 mm) and narrowest medially (L, 1.11 mm), the reverse of what the parapharyngeal border would suggest, while the brainstem is close to isotropic at 0.83 to 0.97 mm, a ratio of 1.17 across directions, as a rigid structure should be. Before any treatment-day image the ordering is the same for every structure measured here: all eight are narrowest superiorly and seven of the eight widest inferiorly, the esophagus alone widest anteriorly, as the high-dose CTV is.

**Table 9.** Agreement of the consensus organ-at-risk contour with the clinician contour on $QACT_{c,2}$, 10 patients, mean ± SD: Dice and ASSD (mm) for the unadapted planning contour on $rTPCT_c$ and for the consensus contours of the three tiers ($\rho$ = 25% for the refined tiers); Δ is the per-patient change relative to $rTPCT_c$.

| Structure | Contour | Dice | ASSD (mm) | Δ Dice | Δ ASSD (mm) |
|---|---|---|---|---|---|
| Oral cavity | $rTPCT_c$ (no adaptation) | 0.900 ± 0.027 | 1.00 ± 0.27 | ref. | ref. |
| | Library-prior $U_L$ consensus | 0.927 ± 0.019 | 0.79 ± 0.24 | +0.027 (9/10, $p < 0.01$) | −0.21 (9/10, $p < 0.05$) |
| | QACT-refined $U_Q$ consensus | 0.943 ± 0.021 | 0.61 ± 0.23 | +0.043 (9/10, $p < 0.01$) | −0.39 (9/10, $p < 0.01$) |
| | Contour-refined $U_C$ consensus | 0.943 ± 0.021 | 0.61 ± 0.24 | +0.043 (9/10, $p < 0.01$) | −0.39 (9/10, $p < 0.01$) |
| Brainstem | $rTPCT_c$ (no adaptation) | 0.864 ± 0.054 | 0.81 ± 0.33 | ref. | ref. |
| | Library-prior $U_L$ consensus | 0.893 ± 0.053 | 0.63 ± 0.30 | +0.030 (7/10, $p = 0.160$) | −0.18 (7/10, $p = 0.105$) |
| | QACT-refined $U_Q$ consensus | 0.950 ± 0.018 | 0.31 ± 0.10 | +0.086 (9/10, $p < 0.01$) | −0.50 (8/10, $p < 0.01$) |
| | Contour-refined $U_C$ consensus | 0.946 ± 0.027 | 0.33 ± 0.16 | +0.083 (8/10, $p < 0.01$) | −0.48 (8/10, $p < 0.01$) |
| Mandible | $rTPCT_c$ (no adaptation) | 0.824 ± 0.045 | 0.81 ± 0.20 | ref. | ref. |
| | Library-prior $U_L$ consensus | 0.875 ± 0.051 | 0.57 ± 0.24 | +0.051 (7/10, $p < 0.05$) | −0.23 (8/10, $p < 0.05$) |
| | QACT-refined $U_Q$ consensus | 0.911 ± 0.042 | 0.42 ± 0.20 | +0.087 (10/10, $p < 0.01$) | −0.39 (9/10, $p < 0.01$) |

| Structure | Contour | Dice | ASSD (mm) | Δ Dice | Δ ASSD (mm) |
|---|---|---|---|---|---|
| | Contour-refined $U_C$ consensus | 0.908 ± 0.045 | 0.43 ± 0.21 | +0.084 (10/10, $p < 0.01$) | −0.38 (9/10, $p < 0.01$) |
| Left parotid | $rTPCT_c$ (no adaptation) | 0.766 ± 0.103 | 1.59 ± 0.87 | ref. | ref. |
| | Library-prior $U_L$ consensus | 0.830 ± 0.090 | 1.08 ± 0.62 | +0.064 (8/10, $p < 0.05$) | −0.52 (8/10, $p < 0.05$) |
| | QACT-refined $U_Q$ consensus | 0.834 ± 0.089 | 1.08 ± 0.68 | +0.069 (9/10, $p < 0.01$) | −0.51 (10/10, $p < 0.01$) |
| | Contour-refined $U_C$ consensus | 0.838 ± 0.081 | 1.04 ± 0.58 | +0.072 (9/10, $p < 0.01$) | −0.55 (10/10, $p < 0.01$) |
| Larynx | $rTPCT_c$ (no adaptation) | 0.794 ± 0.085 | 1.61 ± 0.69 | ref. | ref. |
| | Library-prior $U_L$ consensus | 0.794 ± 0.087 | 1.70 ± 0.76 | +0.000 (5/10, $p = 1.000$) | +0.09 (4/10, $p = 0.492$) |
| | QACT-refined $U_Q$ consensus | 0.826 ± 0.085 | 1.38 ± 0.71 | +0.032 (7/10, $p = 0.065$) | −0.23 (7/10, $p = 0.131$) |
| | Contour-refined $U_C$ consensus | 0.823 ± 0.087 | 1.42 ± 0.75 | +0.029 (6/10, $p = 0.105$) | −0.19 (7/10, $p = 0.193$) |
| Right parotid | $rTPCT_c$ (no adaptation) | 0.751 ± 0.097 | 1.64 ± 0.63 | ref. | ref. |
| | Library-prior $U_L$ consensus | 0.843 ± 0.072 | 0.93 ± 0.36 | +0.092 (9/10, $p < 0.01$) | −0.71 (9/10, $p < 0.01$) |
| | QACT-refined $U_Q$ consensus | 0.838 ± 0.066 | 0.99 ± 0.35 | +0.087 (9/10, $p < 0.01$) | −0.65 (9/10, $p < 0.01$) |
| | Contour-refined $U_C$ consensus | 0.843 ± 0.070 | 0.97 ± 0.40 | +0.092 (9/10, $p < 0.01$) | −0.67 (9/10, $p < 0.01$) |
| Spinal cord | $rTPCT_c$ (no adaptation) | 0.634 ± 0.152 | 1.68 ± 0.98 | ref. | ref. |
| | Library-prior $U_L$ consensus | 0.598 ± 0.131 | 3.04 ± 2.78 | −0.037 (4/10, $p = 0.557$) | +1.36 (3/10, $p = 0.131$) |
| | QACT-refined $U_Q$ consensus | 0.660 ± 0.139 | 1.64 ± 0.94 | +0.026 (7/10, $p = 0.846$) | −0.05 (7/10, $p = 0.695$) |

| Structure | Contour | Dice | ASSD (mm) | Δ Dice | Δ ASSD (mm) |
|---|---|---|---|---|---|
| | Contour-refined $U_C$ consensus | 0.650 ± 0.131 | 1.71 ± 0.93 | +0.015 (6/10, $p = 0.769$) | +0.03 (6/10, $p = 1.000$) |
| Esophagus | $rTPCT_c$ (no adaptation) | 0.506 ± 0.227 | 2.26 ± 1.56 | ref. | ref. |
| | Library-prior $U_L$ consensus | 0.291 ± 0.226 | 7.78 ± 9.11 | −0.215 (2/10, $p < 0.05$) | +5.52 (2/10, $p < 0.05$) |
| | QACT-refined $U_Q$ consensus | 0.475 ± 0.251 | 2.50 ± 1.67 | −0.030 (5/10, $p = 0.695$) | +0.24 (5/10, $p = 0.625$) |
| | Contour-refined $U_C$ consensus | 0.441 ± 0.226 | 2.98 ± 2.38 | −0.065 (4/10, $p = 0.432$) | +0.72 (3/10, $p = 0.160$) |

**Table 10.** Direction-specific 1-$\sigma$ dispersion $\sigma^{loc}{}_{\delta}$ (mm) of the right parotid and brainstem by tier, mean over 10 patients, $\rho = 25\%$ for the refined tiers.

| Structure | Tier | L | R | P | A | S | I |
|---|---|---|---|---|---|---|---|
| Right parotid | Library-prior $U_L$ | 6.00 | 6.86 | 6.18 | 6.29 | 5.01 | 7.62 |
| Right parotid | QACT-refined $U_Q$ | 1.11 | 1.86 | 1.34 | 1.38 | 1.16 | 1.68 |
| Right parotid | Contour-refined $U_C$ | 1.17 | 1.96 | 1.42 | 1.46 | 1.25 | 1.77 |
| Brainstem | Library-prior $U_L$ | 8.61 | 8.56 | 8.34 | 8.83 | 7.96 | 9.46 |
| Brainstem | QACT-refined $U_Q$ | 0.84 | 0.83 | 0.91 | 0.97 | 0.91 | 0.91 |
| Brainstem | Contour-refined $U_C$ | 0.87 | 0.85 | 0.96 | 1.05 | 0.97 | 0.98 |

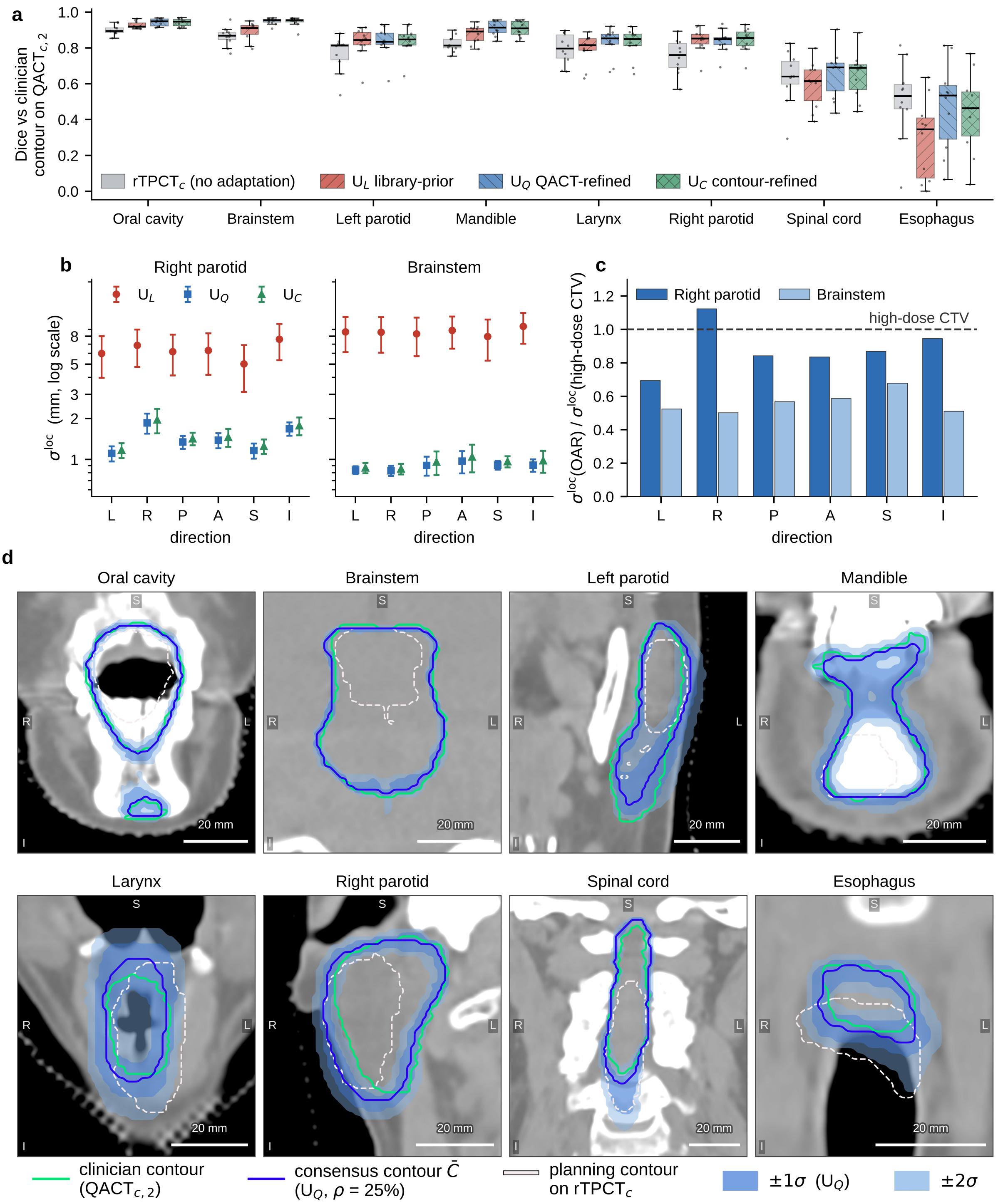

**Figure 5.** Contour forecast and contour uncertainty of organs at risk, 10 patients. (a) Center: Dice of the consensus contour against the clinician contour on QACT$_{c,2}$ for the unadapted planning contour (rTPCT$_c$) and the three tiers, one box per organ, ordered by the unadapted baseline (Table 9). (b) Width: direction-specific 1-$\sigma$ dispersion of the right parotid and brainstem under the three tiers on a common logarithmic axis, $\rho = 25\%$ for the refined tiers (Table 10). (c) Anisotropy of the organ-at-risk dispersion relative to the high-dose CTV, direction by direction; the dashed line is the target itself. (d) The directional band on the treatment-day anatomy for all eight organs, drawn on six of the ten patients: green, clinician contour; dark blue, consensus contour; dashed, planning contour on rTPCT$_c$; shaded, ±1 $\sigma$ and ±2 $\sigma$. The construction is identical to that used for the target.

### 4.4 Validation of the QACT refinement

Three questions are asked of the refined estimate: whether the retention fraction matters, whether the dispersion is ordered correctly, and whether its scale is correct. All validation is against the clinician contour on the patient's second QACT (Section 3.6).

*Retention.* Under QACT refinement the band is insensitive to $\rho$ over most of its range (Supplementary Fig. S2): from $\rho$ = 25% to 90% the direction means move from 1.34–1.78 mm to 1.58–2.09 mm, a change below 0.4 mm in every direction, and the jump to the library-prior width of 4.34 to 6.18 mm on these ten units occurs entirely between 90% and 100%. The library-prior band is therefore wide not because the library disagrees broadly but because roughly one replicate in ten is incompatible with the observed anatomy, and it is the removal of that tenth that the QACT accomplishes. The same structure appears in replicate quality: binned by $s_{img}$ decile, the worst decile has a mean Dice of 0.54 against 0.68 to 0.71 for the other nine, a separation as sharp as the contour-agreement score achieves (0.52 against 0.66 to 0.73). Truncating to $\rho$ = 25% moves the consensus Dice from 0.7266 to 0.7322 and the units beating no adaptation from 7 to 8 of 10; under contour refinement the band widens from 1.64 to 1.98 mm and the Dice falls from 0.744 to 0.728 as $\rho$ rises from 25% to 90%. The 25% operating point is retained.

*Rank.* For each unit and direction the realized error is the root-mean-square mean-centered signed distance from $\bar{C}$ to the clinician contour over the faces of that direction (Figure 6a). The within-unit Spearman correlation between the six $\sigma^{loc}{}_{\delta}$ and the six realized errors has a median of +0.51 under QACT refinement (positive in 8 of 10 units) and +0.40 under contour refinement (8 of 10), against +0.20 (14 of 20) for the library prior. The QACT therefore recovers a per-unit directional ordering that the library alone carries only weakly, and the image is as informative in this respect as an approved contour. In every tier the realized error exceeds the estimated dispersion by a roughly constant factor, which is the scale question.

*Scale.* Area-weighted coverage of the clinician surface by $B_{\kappa}$ is compared with the Gaussian expectation in Table 11 and Figure 6b. Under QACT refinement the band contains 40.9% of the surface at $\kappa$ = 1, 74.6% at 2 and 87.8% at 3, values a Gaussian reaches at 0.54, 1.14 and 1.55 $\sigma$, so by coverage the realized error is 1.75 to 1.94 times the estimated dispersion. Measured instead by the standard deviation of the residual, the ratio is 2.06 (P) to 2.97 (R), about half as large again. Both are correct, and their disagreement is the finding: the residual is not Gaussian. It is negatively skewed in every direction (−0.15 to −0.78), leptokurtic in five of the six (excess kurtosis −0.15 to +3.67), bimodal with a dip at zero, and 6.6 to 21.2% of the surface lies beyond four estimated $\sigma$ (Figure 7 for the QACT-refined tier; library-prior and contour-refined tiers in Supplementary Fig. S3); a standard deviation is inflated by such tails while a central coverage percentile is not, and no single multiplier on $\sigma$ can reconcile the two. Part of the width is a per-unit mean offset, −2.28 mm anteriorly and −1.64 mm on the right against −0.85 to +0.76 mm elsewhere. The same structure appears under contour refinement (ratio 2.07–2.96), so adding the approved contour does not repair the under-dispersion, consistent with Section 4.3. The library-prior band contains its residual comfortably (ratio 0.62–0.78; 80.5% at $\kappa$ = 1 against the Gaussian 68.3%), but this is the counterpart of its being 3.2 to 3.6 times wider, not better calibration. The shortfall is level-dependent: at $\kappa$ = 3 the mid- and low-dose CTVs reach 89.6% and 90.6% while the high-dose CTV reaches 80.4%, so the high-dose residual is heavy-tailed rather than merely wide. Two consequences follow for use. The correspondence between $\kappa$ and coverage must be read from Table 11 rather than from Gaussian tables, as a cohort-level guide rather than a per-patient guarantee; and the estimated $\sigma$ should be used to order and scale directions rather than as the parameter of a normal distribution. At the high-dose level even $\kappa$ = 3 reaches only 80% coverage, so no small multiple of $\sigma$ carries a 90% guarantee there.

*A geometric ceiling.* The flagged structure of Section 3.6 separates a target-definition ceiling from a forecasting failure. A contour of volume V lying inside a ground truth of volume G cannot exceed Dice 2V/(V+G); the QACT-refined consensus contour of this unit (50.84 cc) lies inside its 132.14 cc ground truth to within 0.002 cc, giving a ceiling of 0.556, and the measured value is 0.556. The same patient's low-dose CTV reaches Dice 0.966 with a volume difference of −1.5%, so transport is accurate in this patient and the difficulty is confined to one structure; a uniform 3 mm expansion covers only 54.5% of this target (Section 4.6), which is a failure of the uniform margin rather than of the forecast.

**Table 11.** Area-weighted fraction of the clinician surface within $\pm\kappa\, \sigma^{loc}$ of the consensus contour, $\rho$ = 25% for the refined tiers.

| Band | $U_L$ | $U_Q$ | $U_C$ | Gaussian |
|---|---|---|---|---|
| ±1 $\sigma$, all levels | 80.5% | 40.9% | 42.0% | 68.3% |
| ±2 $\sigma$, all levels | 95.7% | 74.6% | 74.7% | 95.4% |
| ±3 $\sigma$, all levels | 99.0% | 87.8% | 87.8% | 99.7% |
| ±1 $\sigma$, high-dose CTV | 75.9% | 31.5% | 31.1% | 68.3% |
| ±2 $\sigma$, high-dose CTV | 95.8% | 61.7% | 61.0% | 95.4% |
| ±3 $\sigma$, high-dose CTV | 99.4% | 80.4% | 79.8% | 99.7% |

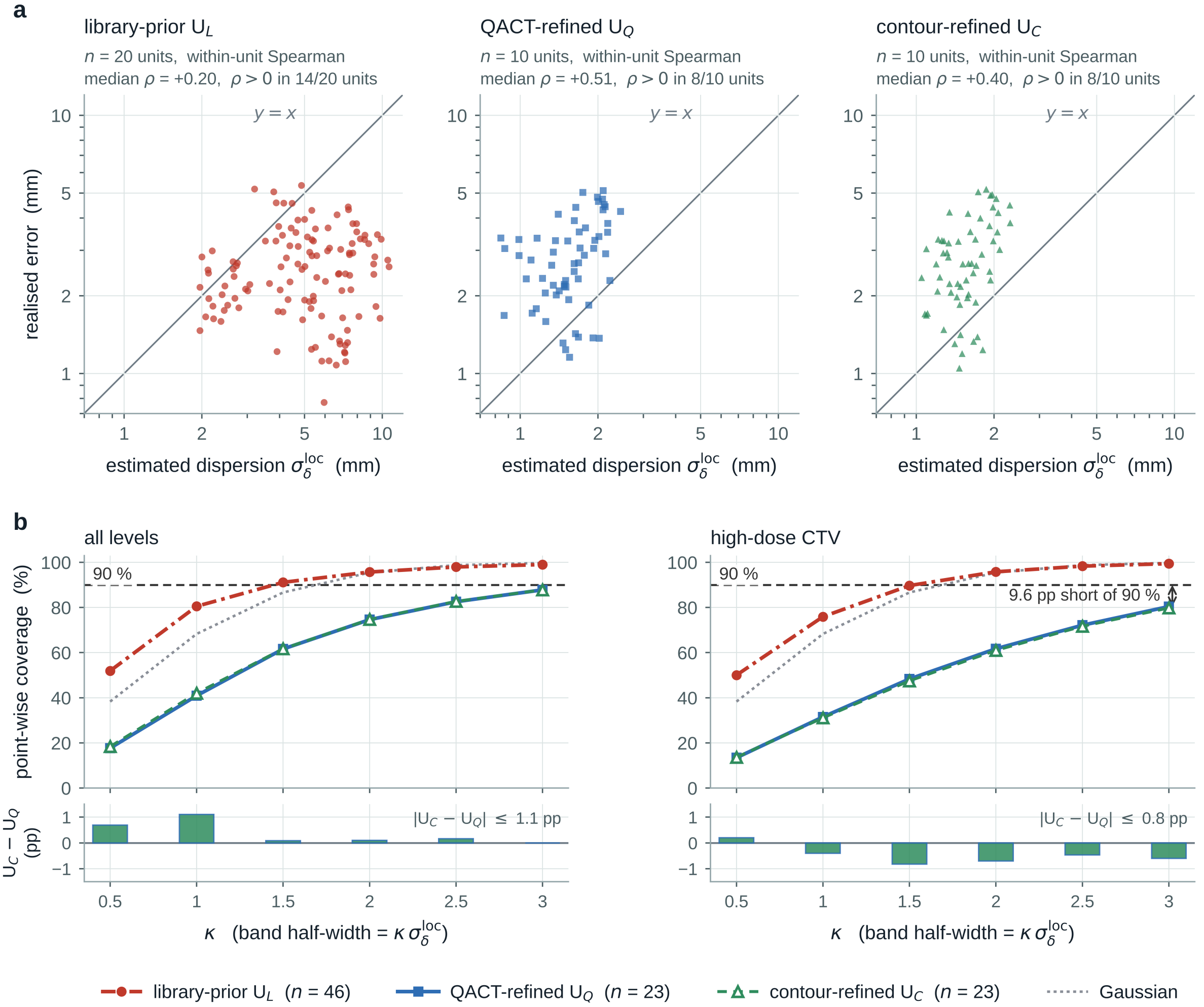

**Figure 6.** Validation of the refined estimate against the clinician contour on $QACT_{c,2}$. (a) Estimated dispersion against realized error, one point per unit and direction, by tier; dashed line, identity; each panel reports the median within-unit Spearman correlation and the number of units in which it is positive, over twenty units for the library prior and ten for the refined tiers, high-dose CTV. (b) Surface coverage against $\kappa$ by tier, with the Gaussian expectation; 46, 23 and 23 units across all dose levels and 20, 10 and 10 at the high-dose level, the library prior having more because it uses no treatment-day information. The values at $\kappa = 1$, 2 and 3 are those of Table 11; the half-steps are computed here. The strip beneath each panel is the contour-refined minus the QACT-refined coverage, which stays within 1.1 percentage points and changes sign between the two groups.

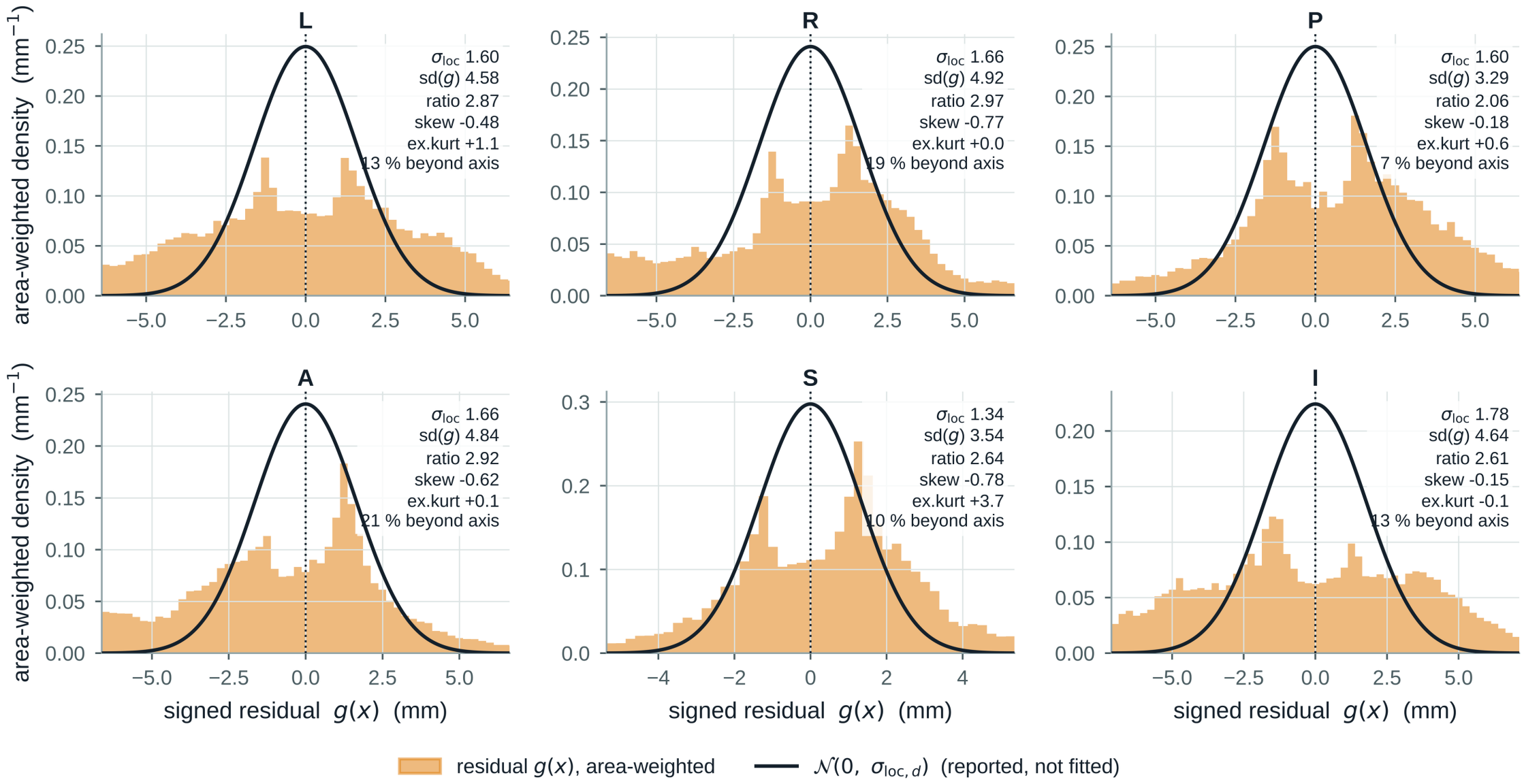

**Figure 7.** Area-weighted distribution of the signed residual by direction, high-dose CTV, with the normal density of the estimated dispersion for reference (not fitted). QACT-refined $U_Q$, $\rho$ = 25%, high-dose CTV, 10 units. On-panel statistics: estimated $\sigma^{loc}$, standard deviation of the residual, their ratio, skewness, excess kurtosis and the fraction of surface beyond the plotted axis. The same distributions for the library-prior and contour-refined tiers are in Supplementary Fig. S3.

### 4.5 Robustness

Table 12 summarizes six experiments. *Reorientation.* Applying the transported displacement without the inverse-Jacobian term, across 80 library–current pairs within the high-dose CTV at a typical deformation of 6.22 mm, produces a median relative error of 36.3% (63.4% at the 90th percentile) and a median directional deviation of 15.2° (30.8°), about 2.26 mm; this exceeds the refined dispersion, so the term is not a second-order correction. *Grid resolution.* Half resolution inflates the dispersion by 16.4% (15.9–23.0%) in all ten patients; sixty spheres of identical radius with sub-voxel jitter, whose true dispersion is zero, measure 1.091 mm at half and 0.540 mm at full resolution, about 0.55 voxel edges added in quadrature, which accounts for 63% of the difference. Convergence at full resolution is therefore not claimed; the reported values are bounds with a known direction. *Score region and form. Restricting the three terms of the image-agreement score to a neighbourhood of the CTV instead of the body mask, or ranking by normalized cross-correlation alone, moves the QACT-refined band by at most 0.054 mm and the center by at most 0.006 Dice, and the volume saving of Section 4.6 by at most 0.43 percentage points, so the conclusions do not depend on the region or form of the score. All three selectors improve the center significantly; the one reported throughout is the most conservative of them (8 of 10 units, $p < 0.05$, against 9 and 10 of 10 for the variants), and it was fixed before the cross-QACT analysis, which is what makes the comparison a check rather than a choice. Library size. Subsampling* the curated twin to n library patients (Supplementary Fig. S4), the library-prior dispersion rises and flattens at 5.42 mm, within 5% of that value from n = 50 (4.10 mm at n = 10, 5.27 at 50, 5.42 at 86); the consensus Dice saturates earlier (0.724 at n = 10, 0.737 at 40, +0.002 by 80); and the standard error of the consensus position falls as $n^{-1/2}$ from 0.72 to 0.32 mm. Additional library patients make the center more stable, not more accurate, and do not narrow the band: the dispersion is a stable property of the library, estimated to within sampling error at the available size. The curated and uncurated twins are compared as a function of library size in Supplementary Fig. S5. *Model form.* Substituting the MIND descriptor for the FlexiCT features (Table 2) changes the selection entirely and the dispersion not at all: the two engines agree on the top-ranked replicate in 0 of 10 patients, the substitute's first choice appearing at ranks 2 to 272 in the primary ordering (Supplementary Table S3), while the directional dispersion is unchanged. The twin therefore contains many near-equivalent replicates rather than a unique optimum, and the uncertainty is a property of the transported deformations, the input class of Table 2, rather than of the engine that ranks them. Curation. The 13 blacklisted sessions (4.4% of the pool) were identified by leave-one-patient-out contour agreement (mean Dice < 0.10); an influence statistic computed without any contour, the change in the consensus when one replicate is removed, reproduces the list with an area under the ROC curve of 0.998, so curation does not require contoured library patients. Sessions that fail on one recipient only are not anticipated by either statistic and are the part of the input uncertainty that the QACT refinement removes.

Two further observations concern selection without the QACT. These use the same-QACT design, since they ask what the library carries offline rather than how the refinement forecasts: twenty patient-date units for the curation folds, eighteen where the flagged unit of Section 4.4 is set aside because its Dice is capped, thirty-six recipient pairs for the transfer test, and ten where the comparison is one per patient. Four independent tests bound what hand-crafted offline scoring of replicate quality can deliver: the library medoid, which is what a recipient would take knowing nothing about that patient, reaches only the 93rd percentile of its own pool, so the value the refinement adds lies in the last seven per cent; replicate quality does transfer between recipients, but weakly, the rank correlation of per-session Dice over the thirty-six pairs of the curated library having a median of 0.26 on the first QACT and 0.29 on the second, positive in 35 and 30 of the pairs, which is seven to eight per cent of shared rank variance and survives the removal of every session that fails outright in either recipient; two feature extractors disagree on every top-ranked replicate at equal accuracy; and retrieval that ignores the QACT, the medoid of the whole pool, fails to beat no adaptation ($p = 0.79$, 7 of 18 units). The same medoid does beat it once the pool is cut to its top quarter, but that cut is made with the treatment-day image and is therefore no longer a QACT-free retrieval. And of the sessions that fail on a held-out patient, the leave-one-patient-out curation anticipates 83% (the mean over twenty held-out patient-date folds, all ten patients included); the remaining 17% are patient-specific incompatibilities that by construction cannot be anticipated from other patients. Both observations are the architectural reason that the refined estimate consumes the patient's QACT, and that what the library supplies before treatment is a distribution rather than a choice.

**Table 12.** Robustness summary of six experiments, high-dose CTV.

| Experiment | Perturbation | Effect on the estimate |
|---|---|---|
| Reorientation | omit inverse-Jacobian term | median 36.3% relative, 15.2° directional error (≈2.26 mm at 6.22 mm deformation); exceeds refined $\sigma$ |
| Grid resolution | full → half | $\sigma$ inflated 16.4% (15.9–23.0%); quantization noise 0.540 → 1.091 mm explains 63% |
| Region of $s_{img}$ | body mask → CTV neighbourhood, or NCC only | band ≤ 0.054 mm, center ≤ 0.006 Dice, volume saving ≤ 0.43 pp |
| Library size | n = 10 → 86 patients | $\sigma$ 4.10 → 5.42 mm, within 5% from n = 50; Dice +0.013; SE of center 0.72 → 0.32 mm ($\propto$ n−1/2) |
| Model form | FlexiCT → MIND | top-1 replicate agreement 0/10; dispersion unchanged |

| Experiment | Perturbation | Effect on the estimate |
|---|---|---|
| Curation | contour-based → influence-based (contour-free) | AUC 0.998 |

### 4.6 Translational example: proton CTV expansion

The uniform expansion is evaluated first because ground truth is available for every unit (Table 13). At 3 mm the expansion covers less than 95% of the treatment-day target in 12 of 20 units (10 of 18 excluding the flagged structure), and the shortfall is not a matter of magnitude: the four worst units have already enlarged the structure by 19 to 74% while missing 10 to 19% of the target, whereas the best-covered units spend volume ratios of 2.3 to 2.6 to reach full coverage. From 3 to 5 mm coverage rises by 0.04 while the volume ratio rises by 28%, so the residual misses are concentrated in particular directions. The same 3 mm produces very different coverage in different patients, which is the direction-dependence and inter-patient spread of Tables 7 and 8 expressed as a margin.

The directional band is compared with the uniform expansion through the coverage–volume curve on all 23 forward structure-QACT units (Figure 8). Inside the clinical window (volume ratio 1.30 to 1.50) the QACT-refined arm covers 1.05 percentage points more of the target than no adaptation at matched volume (median; IQR +0.58 to +1.76; 17 of 18 units; $p < 0.001$), while the library-prior arm is indistinguishable from no adaptation (+0.12 points; 11 of 20; $p = 0.65$). At matched coverage, the refined arm needs 4.1% less volume to reach coverage 0.90 and 3.8% less to reach 0.95 ($p < 0.001$ by unit, $p < 0.01$ clustered), whereas the library-prior arm changes volume by less than 2% at either level and neither change survives clustering by patient. The contour-refined arm gains +1.19 points in the window, so the image alone captures 88% of it. A two-by-two factorial over center and shape at coverage 0.95 on the 21 units for which every cell is defined locates the effect: replacing the planning CTV by the QACT-refined consensus contour saves 5.3% of volume ($p < 0.001$; clustered $p < 0.01$), replacing it by the library-prior consensus contour costs 2.2% without reaching significance ($p = 0.060$), and replacing a uniform by a directional buffer changes volume by 0.2% and is indistinguishable from zero on either center ($p = 0.61$ and 0.94). The volume saving is real but modest and belongs to the center, a better-placed consensus contour obtained from the QACT image alone; the directional shape of the buffer is worth nothing measurable at this operating point, in part because both centers were used in their contracted form (Section 3.5). What the directional estimate supplies is therefore not a smaller expansion but a differently informed one, patient- and direction-specific with an empirically known coverage for each $\kappa$. The representable-variance ratio r of Section 3.5 is 0.247 across the cohort: a per-axis constant field, which is what a conventional recipe is within each direction sector, can represent at most one quarter of the contour-uncertainty variance, and this is a conservative bound.

**Table 13.** Uniform expansion of the planning high-dose CTV on $rTPCT_c$: volumetric coverage of the treatment-day target and volume ratio, 20 units.

| r (mm) | Coverage, mean | Coverage, min | Units < 95% | Volume ratio, mean |
|---|---|---|---|---|
| 2 | 0.861 | 0.481 | 13/20 | 1.36 |
| 3 | 0.898 | 0.545 | 12/20 | 1.58 |
| 4 | 0.919 | 0.591 | 9/20 | 1.76 |
| 5 | 0.939 | 0.649 | 5/20 | 2.03 |

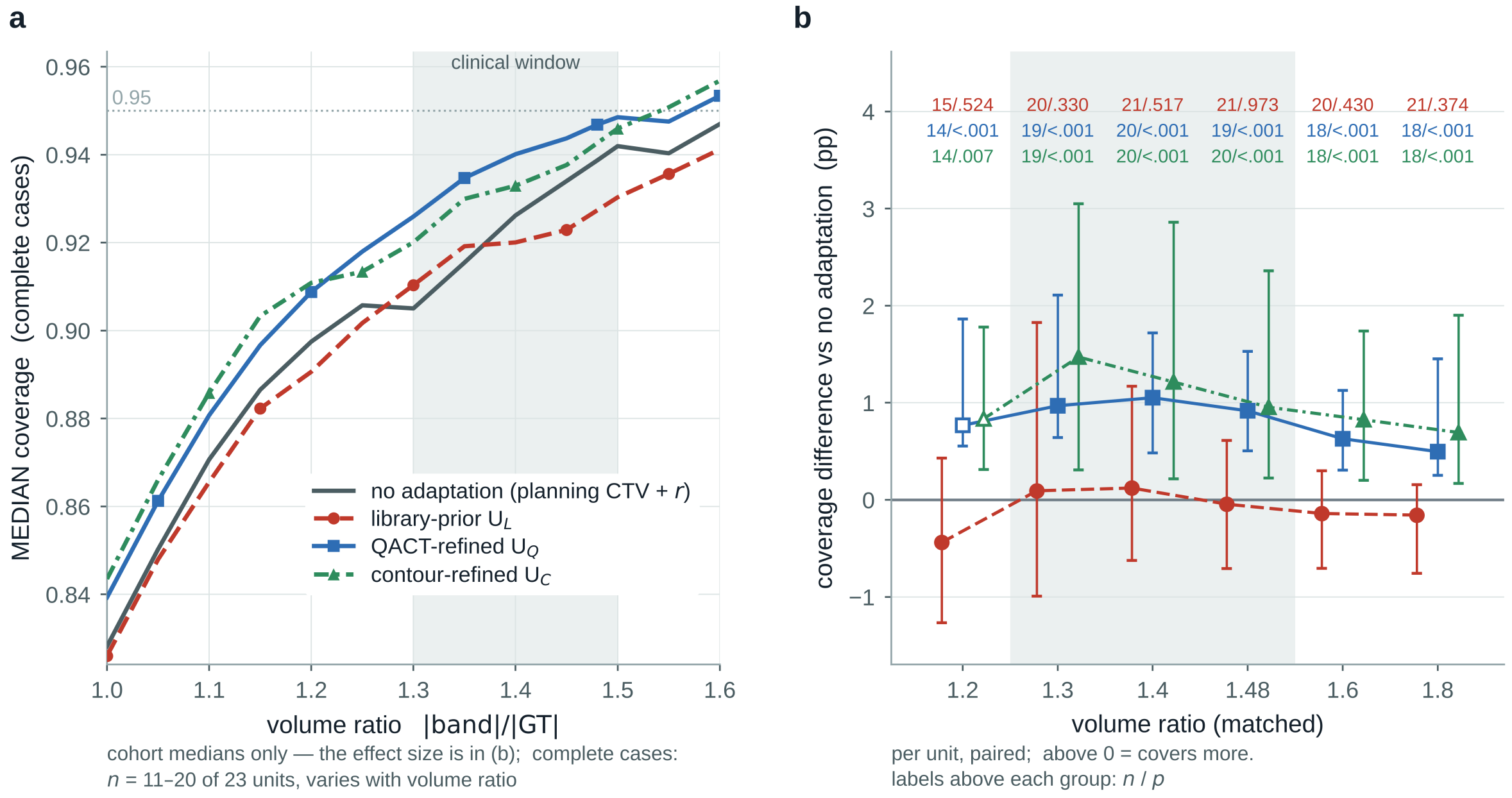

**Figure 8.** Coverage of the treatment-day target against irradiated volume, 23 forward structure-QACT units. (a) Cohort medians on a volume-ratio grid with the clinical window shaded; complete cases only, so each volume ratio rests on 11 to 20 of the 23 units. (b) Per-unit paired difference of each arm against its own unadapted baseline at matched volume: median, IQR, paired units and Wilcoxon p.

## 5. Discussion

The results separate three contributions to the forecast, and the separation is the main finding. The shape of the contour uncertainty belongs to the library and to the current patient's anatomy: the library-prior band is anisotropic, tightest superiorly and widest inferiorly, and the six directions keep the same rank order under all three tiers (Table 7, Figure 3). The scale belongs to the patient's own QACT: refinement on $\mathrm{QACT}_{c,1}$ narrows every direction by a factor of 3.2 to 3.6 without any contour being drawn on it, and the retention sweep locates the mechanism, since the band is flat from $\rho = 25\%$ to 90% and jumps to the library-prior width only between 90% and 100% (Section 4.4). The library is wide not because previously treated patients disagree broadly about how a head and neck changes, but because roughly one replicate in ten is incompatible with the anatomy this patient actually presents, and the image is what identifies that tenth. The center belongs to whatever same-patient information is most specific: the QACT image improves the consensus contour by +0.015 Dice and a physician-approved contour on it by a further +0.012, while leaving the width within 0.06 mm of the image-only value in every direction (Tables 6 and 7). This division of labour is what makes the twin useful at two moments: at planning the anisotropic shape is already available, which is exactly what a uniform expansion discards, and during treatment the first QACT converts it into a band whose width has been set by observed anatomy, without synthesis, segmentation or optimization being repeated on the day.

The quantity estimated is input uncertainty in the sense of Table 2, and two robustness results confirm this. Exchanging the registration feature extractor changes the top-ranked replicate in every patient but leaves the dispersion unchanged, and enlarging the library from 10 to 86 patients stabilises the center as $n^{-1/2}$ without narrowing the band (Table 12): the dispersion is a property of the transported deformations, estimated to within sampling error at the available size, and only data about the current patient reduces it. Three properties follow from the construction: every replicate is a diffeomorphic warp of the patient's own anatomy by a deformation that occurred, so no implausible shape can be synthesized; the output is a length in millimeters resolved by direction rather than an entropy or a logit variance; and the ensemble can be inspected member by member, so an implausible replicate can be removed and the band recomputed.

The estimate is rank-informative but not Gaussian-calibrated, and Table 11 and Figure 6 show why the two must be kept apart. Directions with larger estimated $\sigma$ are those in which the consensus contour is in fact further from the clinician contour (within-unit Spearman +0.51 under QACT refinement against +0.20 for the library alone), so the QACT recovers a per-patient directional ordering that the library carries only weakly. The absolute scale is another matter. The realized residual is negatively skewed, leptokurtic and bimodal, with 7 to 21% of the surface beyond four estimated $\sigma$, so the coverage route and the standard-deviation route to a calibration factor disagree by about 1.5, and no single multiplier reconciles them. Adding the approved contour does not repair this (Table 11), which is consistent with the width being fixed by the image. The consequence for use is direct: $\kappa$ should be read against the empirical coverage of Table 11 rather than against a normal table, as a cohort-level guide rather than a per-patient guarantee, and at the high-dose level, where the residual is heavy-tailed rather than merely wide, even $\kappa = 3$ reaches only 80% surface coverage. The dispersion is best used to order and scale directions, which it does well, rather than as the parameter of a distribution, which it is not.

Two negative results bound the claims. The library does not improve the point forecast: the library-prior consensus contour is indistinguishable from the planning contour (Table 6), so what the population supplies before treatment is the distribution, not a better center. And in the translational example the directional shape of the buffer does not save volume: the QACT-refined arm does irradiate about 4% less volume than a uniform expansion at matched coverage (Figure 8), but the factorial places that saving in the center, a consensus contour better placed by the QACT image, while replacing a uniform by a directional buffer moves volume by 0.1 to 0.2% on either center. Part of the reason is methodological: averaging signed distance fields contracts the consensus contour, more so for the wide library-prior set (0.86 of the target volume, all dose levels) than for the refined set (0.96), so the band must first restore a deficit that the aggregation created, and the library-prior arm falls below the unadapted baseline for that reason alone. What the directional estimate defends is therefore not a smaller expansion but a differently informed one. A uniform 3 mm expansion covers less than 95% of the treatment-day target in 12 of 20 units while having already enlarged the worst of them by 19 to 74% (Table 13); the per-patient dispersion varies two- to three-fold within a single direction (Table 8); and a per-axis constant field can represent at most a quarter of the contour-uncertainty variance ($r = 0.247$). Patient- and direction-specific is the claim the data support; smaller is not.

Several limitations bound the work. The refinement is validated with the patient's first QACT informing a forecast of the second, and the two are typically one to two weeks apart, so the demonstrated forecasting horizon is short; this is a limitation of the retrospective data rather than of the method, and an image acquired on the first treatment day, whether an early QACT or a synthetic CT derived from CBCT, would extend it. The evaluation cohort is ten patients from one institution, the refined tiers have ten validation units, and comparisons that use no treatment-day information have an effective sample size of ten patients, which is why clustered p-values are reported alongside per-unit ones and why the two disagree for the library-prior arm in Figure 8. Uncertainty is reported for contours only; the predicted images are characterized for fidelity (Figure 2) but their voxel-wise dispersion is not quantified. Organ-at-risk results are reported for the eight structures that carry clinician contours in the aligned ground truth and the directional dispersion for two of them, and the two thin tubular structures gain nothing from any tier, so the estimate is demonstrated for compact organs and not for elongated ones. The curation threshold was chosen empirically and not swept, so the sensitivity of the estimate to a poorly chosen value is unknown, and the curation would have to be repeated on another institution's library. The dispersion at full resolution has not been shown to converge with grid spacing, and the representable-variance bound of 0.247 is conservative; both should be read as bounds with a known direction. No dose is calculated, so the clinical value of the geometric result remains an inference. Future work will focus on expanding the imaging cohort, in library size and in institutions, to widen the domain over which the framework applies.

## 6. Conclusion

We have presented a digital-twin framework that forecasts the treatment-day anatomy of a head-and-neck proton patient as an ensemble of predicted CTs with propagated contours, and that reports the uncertainty of the forecast contours in millimeters, resolved by direction. The twin is a library of previously treated patients made patient-specific by a two-step foundation-model registration, so that every member is a deformation that occurred in a treated patient applied to the current patient's own anatomy. The uncertainty so obtained is uncertainty in the input to the forecast, which library patient the current patient will follow, rather than in the form or parameters of a model: it is unchanged when the registration engine is exchanged, it does not narrow as the library grows, and it narrows only when data about the current patient arrive. The library supplies the anisotropic shape of the uncertainty before treatment; the patient's first QACT sets its scale, narrowing every direction by more than a factor of three without a contour being drawn; a physician-approved contour improves the center and not the width. The estimate orders directions correctly but is not Gaussian-calibrated, and its coverage must be read empirically. The clinical target volume expansion worked out here is one application of a general construction that applies without modification to any delineated structure.

### CRediT authorship contribution statement

**Yizhou Wu:** Conceptualization, Data curation, Formal analysis, Investigation, Software, Validation, Visualization, Writing – original draft. **Jie Ding:** Resources, Validation, Writing – review & editing. **Justin Roper:** Resources, Validation, Writing – review & editing. **Minglei Kang:** Resources, Validation, Writing – review & editing. **Yuheng Li:** Resources, Software. **Sibo Tian:** Data curation, Investigation, Validation, Writing – review & editing. **David S. Yu:** Conceptualization, Resources, Writing – review & editing. **Xiaofeng Yang:** Conceptualization, Funding acquisition, Supervision, Writing – review & editing. **Chih-Wei Chang:** Conceptualization, Funding acquisition, Methodology, Project administration, Supervision, Writing – review & editing.

### Declaration of Competing Interest

The authors declare that they have no known competing financial interests or personal relationships that could have appeared to influence the work reported in this paper.

### Data availability

The imaging data supporting this study are not publicly available because they contain protected health information. Derived quantities and analysis code are available from the corresponding author on reasonable request.

### Ethics statement

Emory IRB review board approval was obtained, and informed consent was not required for this Health Insurance Portability and Accountability Act (HIPAA) compliant retrospective analysis. The research was conducted in accordance with the principles embodied in the Declaration of Helsinki and in accordance with local statutory requirements.

### Funding

This research is supported in part by the National Institutes of Health under Award Number R01DE033512, and R01CA272991.

# Supplementary material

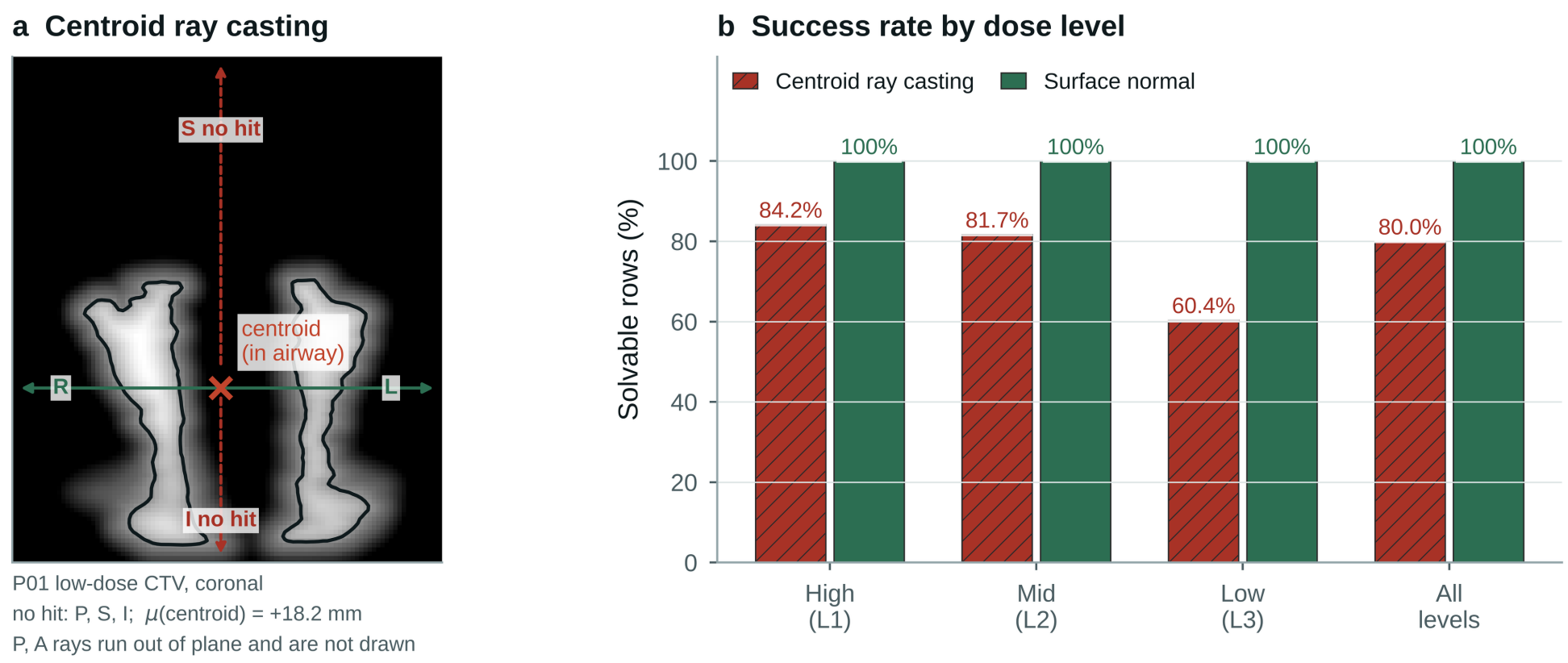

**Figure S1.** Why the surface-normal parameterization is needed. (a) Centroid ray casting on a collar-shaped low-dose CTV: the centroid falls in the airway and the posterior, superior and inferior rays never meet the surface. (b) Fraction of structure-by-direction combinations solvable by ray casting, by dose level, against 100% for the surface-normal parameterization.

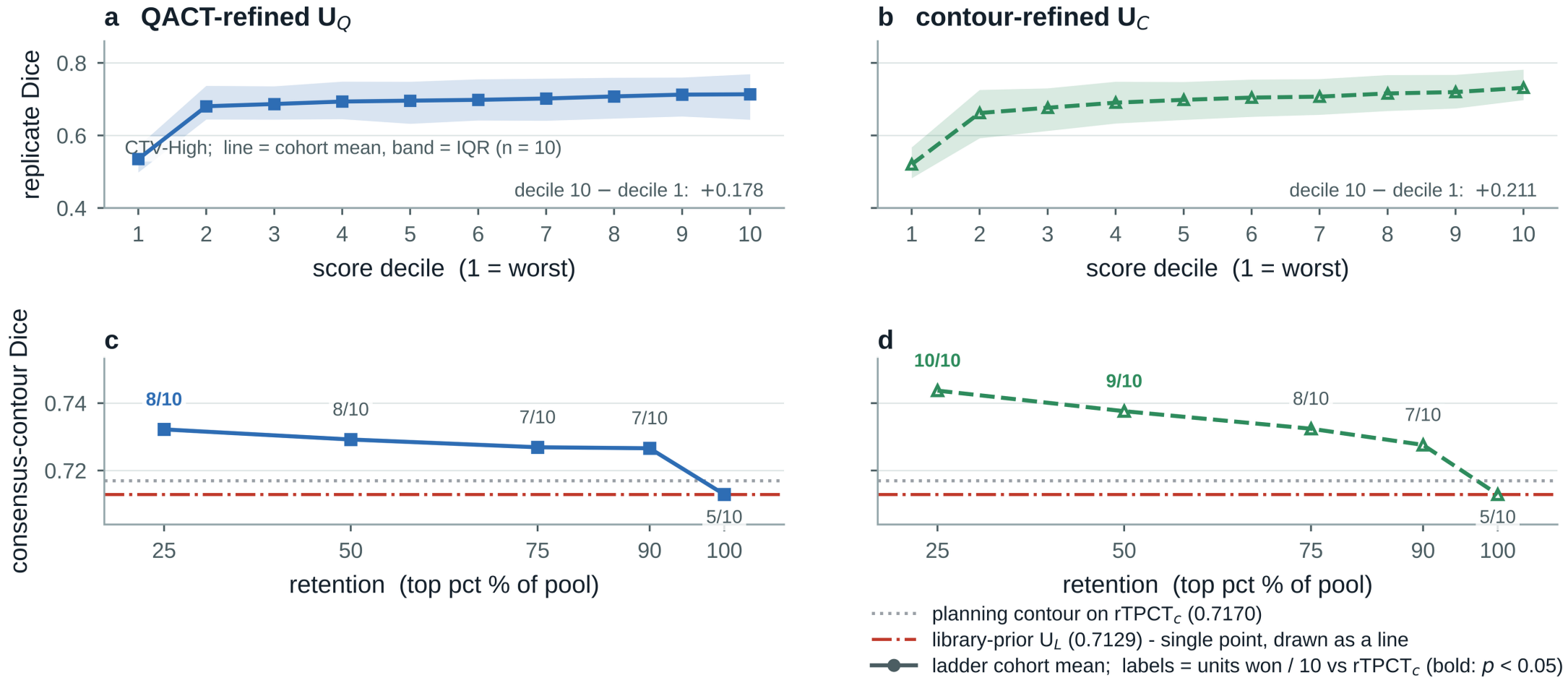

**Figure S2.** Retention sweep, high-dose CTV. Top: replicate Dice by score decile under the image-agreement (left) and contour-agreement (right) scores. Bottom: consensus-contour Dice against retention fraction $\rho$, with the planning contour on $rTPCT_c$ and the library-prior value as reference lines.

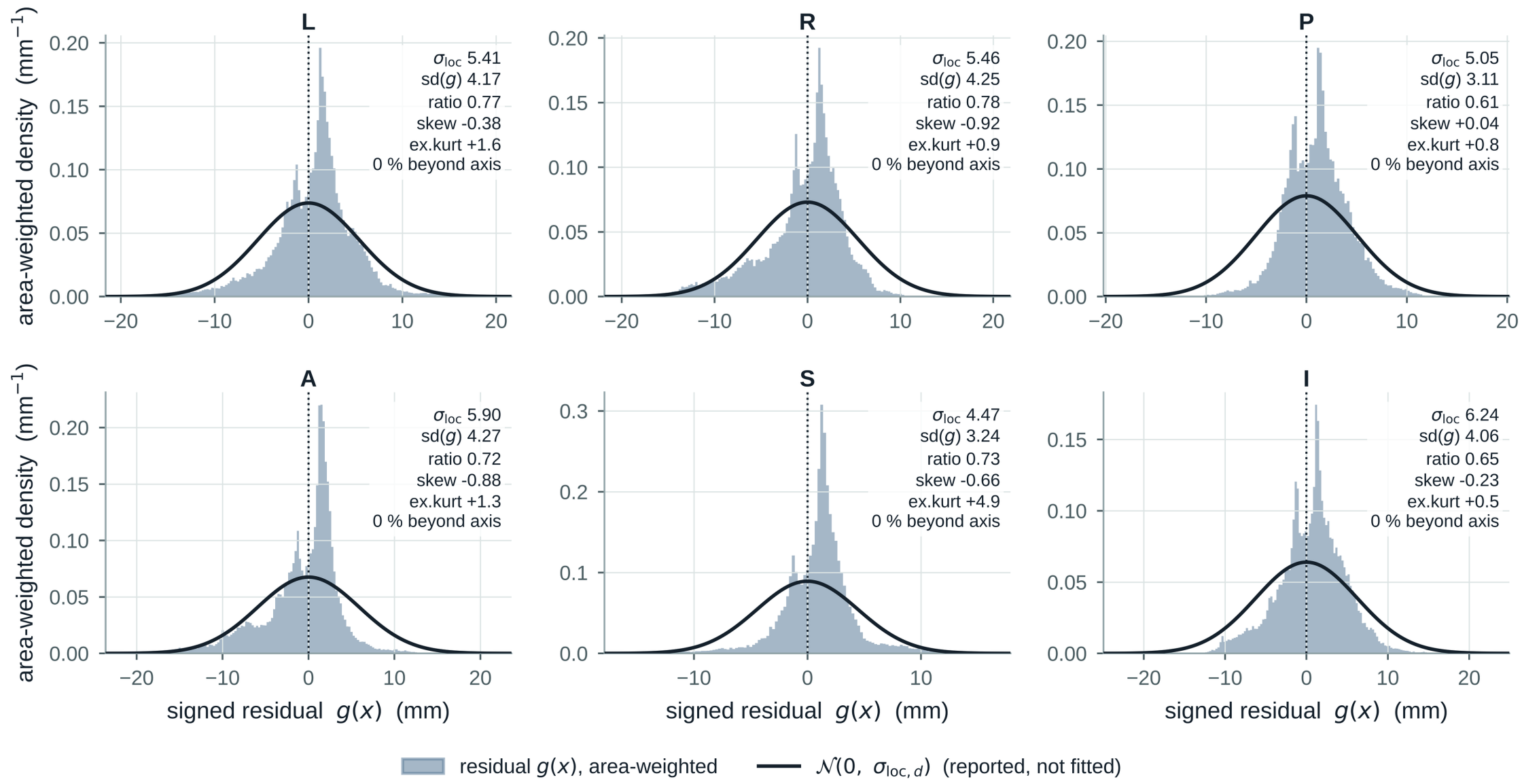

**Figure S3.** Area-weighted distribution of the signed residual by direction, high-dose CTV, with the normal density of the estimated dispersion for reference (not fitted); companion to Figure 7. (a) Library-prior $U_L$.

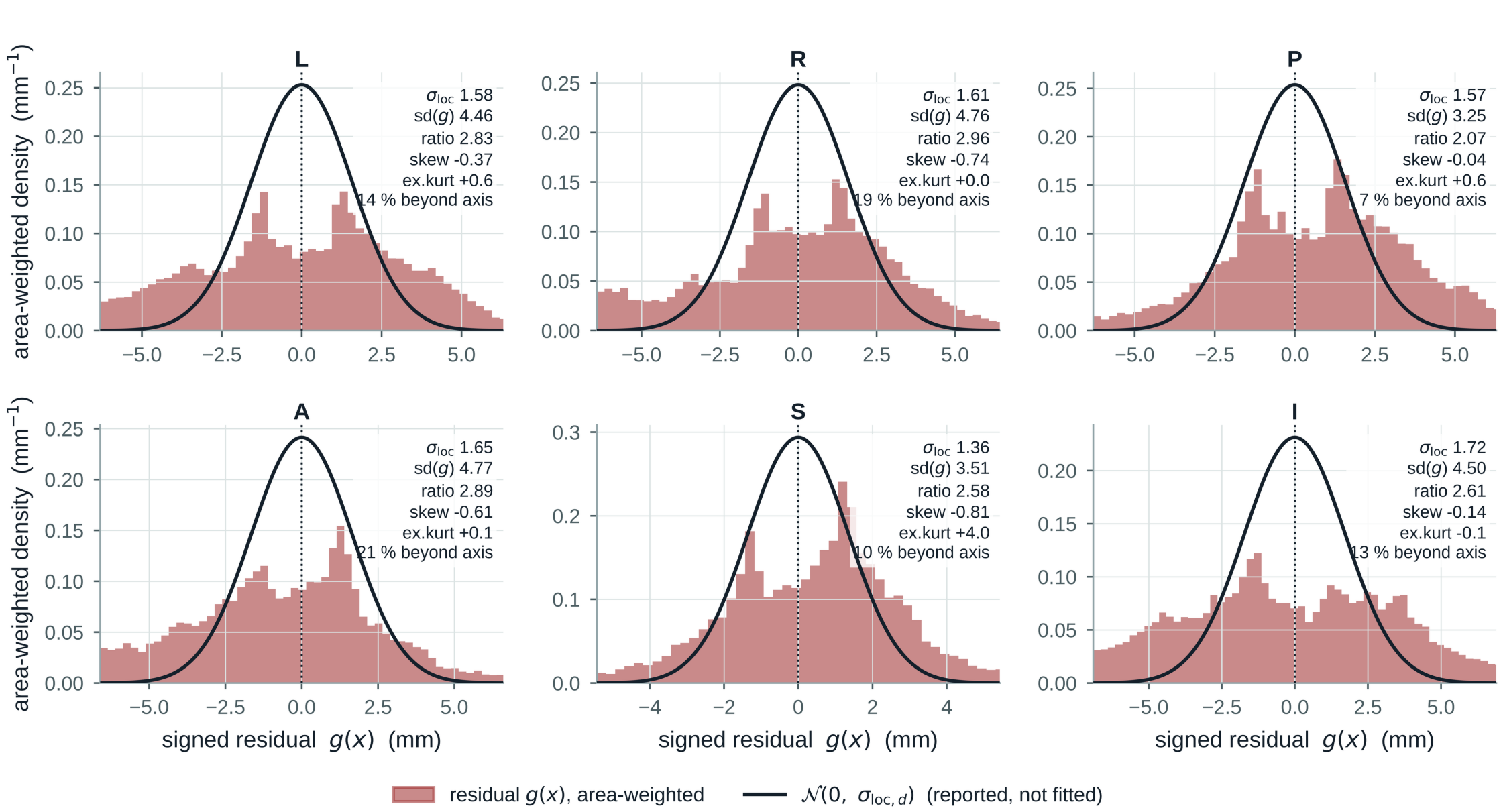

**Figure S3** (continued). (b) Contour-refined $U_C$, $\rho$ = 25%.

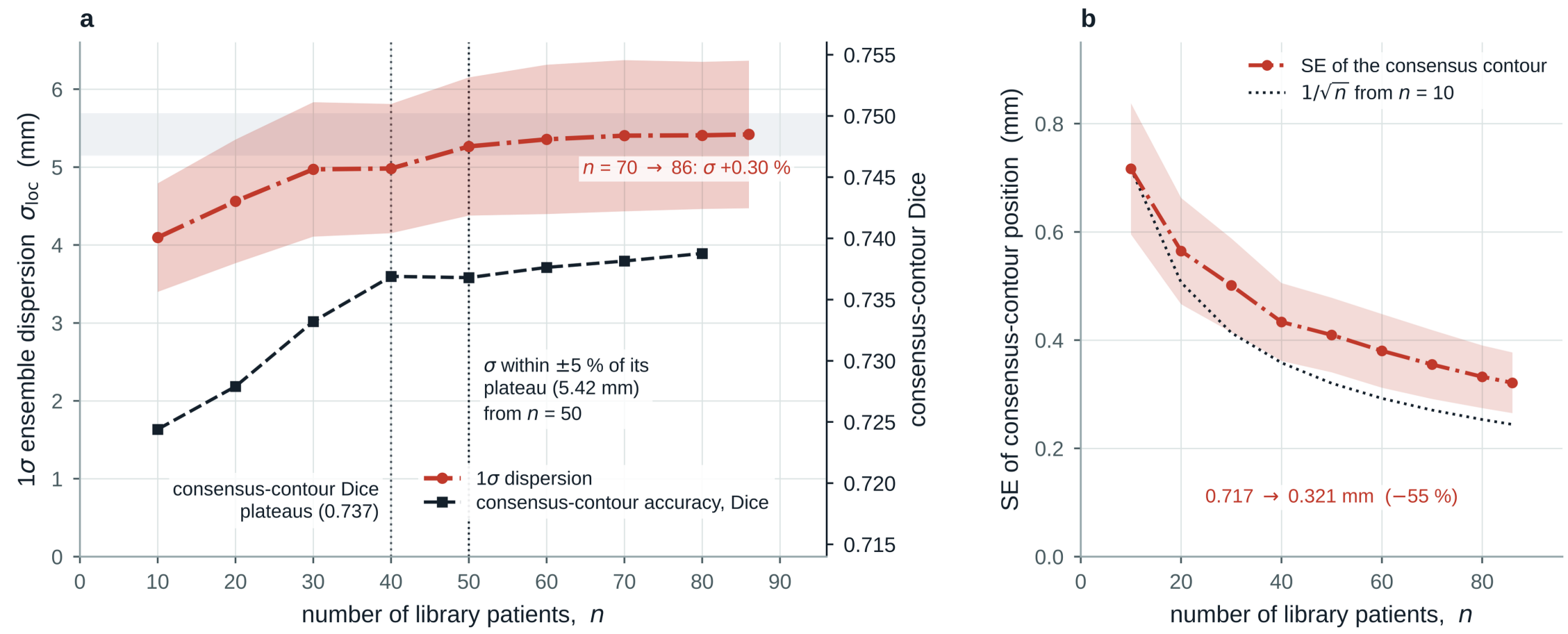

**Figure S4**. Effect of library size, high-dose CTV, 20 random draws per size; cohort means with 95% confidence interval. (a) Library-prior dispersion (left axis) and consensus-contour Dice (right axis) against the number of library patients. (b) Standard error of the consensus-contour position, falling as n^−1/2.

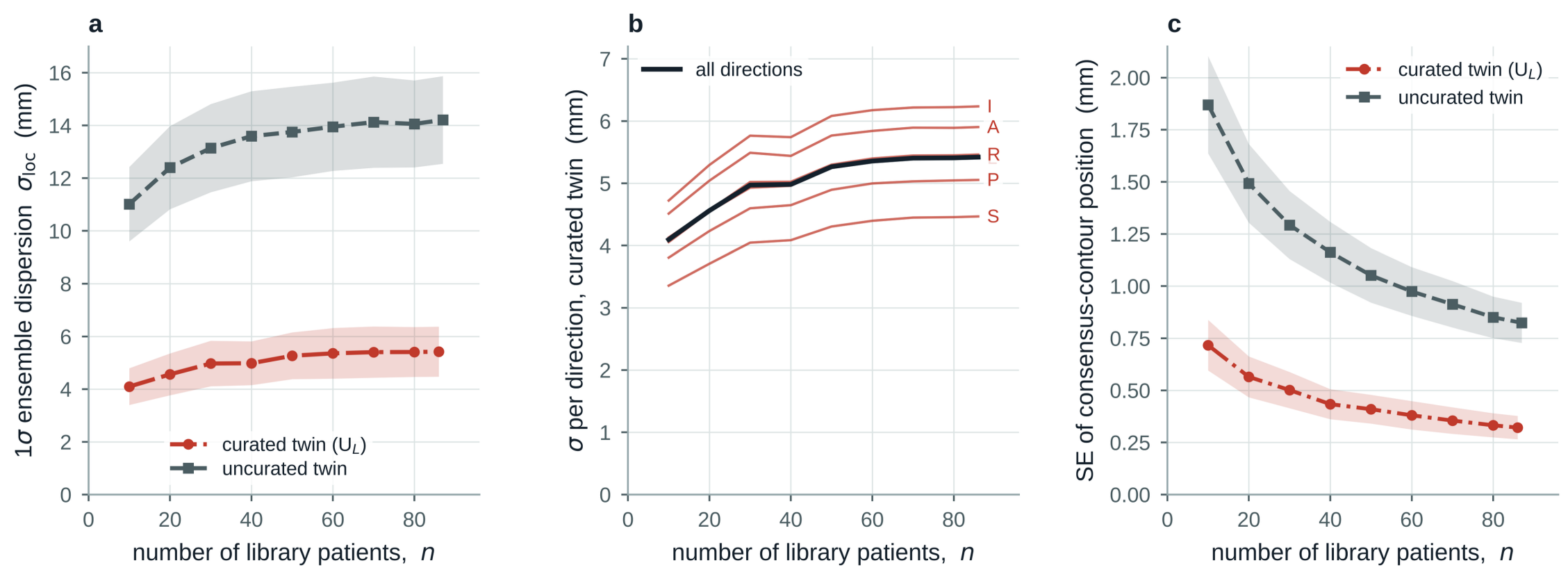

**Figure S5.** Curated against uncurated twin as a function of library size: (a) dispersion, (b) per-direction dispersion of the curated twin, (c) standard error of the consensus-contour position.

**Table S1.** Image agreement of each retained set with the treatment-day QACT. Cross-QACT design: the retained set is chosen on $QACT_{c,1}$ and scored against $QACT_{c,2}$. Values are means over the ten forward units. NCC and SSIM: higher is better; MAE: lower is better. Agreement rises with selectivity, from the whole curated pool through the top quarter to the single highest-ranked replicate.

| Arm | n | NCC | SSIM | MAE |
|---|---|---|---|---|
| $rTPCT_c$ (no adaptation) | 10 | 0.9291 | 0.7953 | 0.0241 |
| Library-prior $U_L$ (whole curated pool) | 10 | 0.8945 | 0.7545 | 0.0307 |
| QACT-refined $U_Q$, rank-1 | 10 | 0.9311 | 0.8031 | 0.0239 |
| QACT-refined $U_Q$, top 25% | 10 | 0.9222 | 0.7841 | 0.0263 |
| Contour-refined $U_C$, rank-1 | 10 | 0.9211 | 0.7830 | 0.0261 |

| Arm | n | NCC | SSIM | MAE |
|---|---|---|---|---|
| Contour-refined $U_C$, top 25% | 10 | 0.9170 | 0.7766 | 0.0273 |

**Table S2.** Direction-specific dispersion and point-wise calibration by dose level. $\sigma_{loc}$ is summarized over units in mm; coverage is the area-weighted fraction of the clinician surface inside $B_\kappa$, in per cent, pooled over all facets of all units at that level. The library prior applies no ranking and therefore uses both QACTs (n = 20 / 20 / 6 for L1 / L2 / L3); the refined tiers are cross-QACT forward units (n = 10 / 10 / 3). L1 repeats the main text for reference.

| Level | Tier | n | Dir. | Mean | P25 | P75 | Min | Max | ±1$\sigma$ | ±2$\sigma$ | ±3$\sigma$ |
|---|---|---|---|---|---|---|---|---|---|---|---|
| L1 | $U_L$ | 20 | L | 5.41 | 3.99 | 7.14 | 1.97 | 9.90 | 75.9 | 95.8 | 99.4 |
| L1 | $U_L$ | 20 | R | 5.46 | 3.98 | 7.06 | 2.37 | 9.80 | 75.9 | 95.8 | 99.4 |
| L1 | $U_L$ | 20 | P | 5.05 | 3.96 | 6.77 | 2.11 | 9.35 | 75.9 | 95.8 | 99.4 |
| L1 | $U_L$ | 20 | A | 5.90 | 4.36 | 7.41 | 2.52 | 10.61 | 75.9 | 95.8 | 99.4 |
| L1 | $U_L$ | 20 | S | 4.47 | 3.16 | 5.52 | 1.97 | 9.57 | 75.9 | 95.8 | 99.4 |
| L1 | $U_L$ | 20 | I | 6.24 | 4.78 | 7.79 | 2.65 | 10.49 | 75.9 | 95.8 | 99.4 |
| L2 | $U_L$ | 20 | L | 4.25 | 3.40 | 4.93 | 2.47 | 8.57 | 80.2 | 95.0 | 98.7 |
| L2 | $U_L$ | 20 | R | 4.32 | 3.60 | 4.86 | 2.38 | 8.31 | 80.2 | 95.0 | 98.7 |
| L2 | $U_L$ | 20 | P | 4.68 | 3.97 | 5.49 | 2.51 | 8.00 | 80.2 | 95.0 | 98.7 |
| L2 | $U_L$ | 20 | A | 5.27 | 4.31 | 5.97 | 2.73 | 9.73 | 80.2 | 95.0 | 98.7 |
| L2 | $U_L$ | 20 | S | 3.63 | 2.58 | 3.98 | 1.98 | 7.66 | 80.2 | 95.0 | 98.7 |
| L2 | $U_L$ | 20 | I | 5.97 | 5.40 | 6.87 | 3.04 | 9.70 | 80.2 | 95.0 | 98.7 |
| L3 | $U_L$ | 6 | L | 4.21 | 4.06 | 4.33 | 3.97 | 4.56 | 87.3 | 97.6 | 99.6 |
| L3 | $U_L$ | 6 | R | 4.20 | 4.18 | 4.25 | 4.02 | 4.29 | 87.3 | 97.6 | 99.6 |
| L3 | $U_L$ | 6 | P | 4.66 | 4.24 | 5.09 | 4.11 | 5.61 | 87.3 | 97.6 | 99.6 |
| L3 | $U_L$ | 6 | A | 5.41 | 4.78 | 5.92 | 4.74 | 6.76 | 87.3 | 97.6 | 99.6 |
| L3 | $U_L$ | 6 | S | 3.22 | 2.70 | 3.58 | 2.53 | 4.08 | 87.3 | 97.6 | 99.6 |
| L3 | $U_L$ | 6 | I | 5.59 | 4.95 | 5.94 | 4.68 | 6.58 | 87.3 | 97.6 | 99.6 |
| L1 | $U_Q$ | 10 | L | 1.60 | 1.38 | 1.92 | 0.84 | 2.14 | 31.5 | 61.7 | 80.4 |
| L1 | $U_Q$ | 10 | R | 1.66 | 1.36 | 2.02 | 0.87 | 2.10 | 31.5 | 61.7 | 80.4 |
| L1 | $U_Q$ | 10 | P | 1.60 | 1.40 | 1.70 | 0.87 | 2.19 | 31.5 | 61.7 | 80.4 |
| L1 | $U_Q$ | 10 | A | 1.66 | 1.43 | 1.88 | 1.10 | 2.23 | 31.5 | 61.7 | 80.4 |
| L1 | $U_Q$ | 10 | S | 1.34 | 1.13 | 1.49 | 0.99 | 1.77 | 31.5 | 61.7 | 80.4 |
| L1 | $U_Q$ | 10 | I | 1.78 | 1.55 | 2.05 | 0.99 | 2.45 | 31.5 | 61.7 | 80.4 |
| L2 | $U_Q$ | 10 | L | 1.67 | 1.51 | 1.95 | 0.83 | 2.34 | 43.9 | 78.4 | 89.6 |
| L2 | $U_Q$ | 10 | R | 1.78 | 1.68 | 2.02 | 0.89 | 2.33 | 43.9 | 78.4 | 89.6 |
| L2 | $U_Q$ | 10 | P | 1.81 | 1.58 | 2.17 | 0.89 | 2.51 | 43.9 | 78.4 | 89.6 |
| L2 | $U_Q$ | 10 | A | 1.92 | 1.71 | 2.18 | 1.11 | 2.71 | 43.9 | 78.4 | 89.6 |

| Level | Tier | n | Dir. | Mean | P25 | P75 | Min | Max | ±1$\sigma$ | ±2$\sigma$ | ±3$\sigma$ |
|---|---|---|---|---|---|---|---|---|---|---|---|
| L2 | $U_Q$ | 10 | S | 1.52 | 1.25 | 1.85 | 0.92 | 2.17 | 43.9 | 78.4 | 89.6 |
| L2 | $U_Q$ | 10 | I | 1.96 | 1.70 | 2.18 | 1.16 | 2.84 | 43.9 | 78.4 | 89.6 |
| L3 | $U_Q$ | 3 | L | 1.75 | 1.67 | 1.81 | 1.61 | 1.90 | 42.4 | 77.8 | 90.6 |
| L3 | $U_Q$ | 3 | R | 1.92 | 1.73 | 2.06 | 1.64 | 2.30 | 42.4 | 77.8 | 90.6 |
| L3 | $U_Q$ | 3 | P | 1.85 | 1.79 | 1.94 | 1.66 | 1.97 | 42.4 | 77.8 | 90.6 |
| L3 | $U_Q$ | 3 | A | 1.91 | 1.80 | 2.10 | 1.54 | 2.12 | 42.4 | 77.8 | 90.6 |
| L3 | $U_Q$ | 3 | S | 1.30 | 1.13 | 1.47 | 0.96 | 1.65 | 42.4 | 77.8 | 90.6 |
| L3 | $U_Q$ | 3 | I | 1.86 | 1.74 | 2.00 | 1.57 | 2.10 | 42.4 | 77.8 | 90.6 |
| L1 | $U_C$ | 10 | L | 1.58 | 1.35 | 1.74 | 1.19 | 2.30 | 31.1 | 61.0 | 79.8 |
| L1 | $U_C$ | 10 | R | 1.61 | 1.38 | 1.83 | 1.09 | 2.31 | 31.1 | 61.0 | 79.8 |
| L1 | $U_C$ | 10 | P | 1.57 | 1.37 | 1.67 | 1.09 | 2.06 | 31.1 | 61.0 | 79.8 |
| L1 | $U_C$ | 10 | A | 1.65 | 1.40 | 1.91 | 1.21 | 1.99 | 31.1 | 61.0 | 79.8 |
| L1 | $U_C$ | 10 | S | 1.36 | 1.24 | 1.47 | 1.05 | 1.74 | 31.1 | 61.0 | 79.8 |
| L1 | $U_C$ | 10 | I | 1.72 | 1.50 | 2.01 | 1.27 | 2.10 | 31.1 | 61.0 | 79.8 |
| L2 | $U_C$ | 10 | L | 1.68 | 1.48 | 1.93 | 1.04 | 2.25 | 44.0 | 77.9 | 89.6 |
| L2 | $U_C$ | 10 | R | 1.79 | 1.72 | 2.04 | 1.11 | 2.16 | 44.0 | 77.9 | 89.6 |
| L2 | $U_C$ | 10 | P | 1.88 | 1.68 | 2.23 | 1.06 | 2.56 | 44.0 | 77.9 | 89.6 |
| L2 | $U_C$ | 10 | A | 2.00 | 1.76 | 2.30 | 1.37 | 2.87 | 44.0 | 77.9 | 89.6 |
| L2 | $U_C$ | 10 | S | 1.55 | 1.25 | 1.87 | 1.03 | 2.16 | 44.0 | 77.9 | 89.6 |
| L2 | $U_C$ | 10 | I | 1.99 | 1.63 | 2.22 | 1.38 | 2.78 | 44.0 | 77.9 | 89.6 |
| L3 | $U_C$ | 3 | L | 1.90 | 1.88 | 1.92 | 1.86 | 1.95 | 48.7 | 80.8 | 91.4 |
| L3 | $U_C$ | 3 | R | 2.02 | 1.97 | 2.05 | 1.96 | 2.11 | 48.7 | 80.8 | 91.4 |
| L3 | $U_C$ | 3 | P | 1.97 | 1.85 | 2.04 | 1.83 | 2.21 | 48.7 | 80.8 | 91.4 |
| L3 | $U_C$ | 3 | A | 2.10 | 1.93 | 2.22 | 1.86 | 2.44 | 48.7 | 80.8 | 91.4 |
| L3 | $U_C$ | 3 | S | 1.39 | 1.23 | 1.51 | 1.15 | 1.72 | 48.7 | 80.8 | 91.4 |
| L3 | $U_C$ | 3 | I | 1.92 | 1.79 | 2.02 | 1.72 | 2.19 | 48.7 | 80.8 | 91.4 |

**Table S3**. Engine substitution. For each patient, the highest-ranked library member chosen by FlexiCT and by MIND on $QACT_{c,2}$, and the rank that each engine's choice receives under the other engine's ordering, over the replicates that both engines scored. Pool is the number of library observations available to that patient once the patient's own observations and any failed transports are removed; scored by both is how many of those carry a valid score from each engine, and is lower than the pool only for P03 and P08, where some transports yielded no FlexiCT score. Library patients are coded L01–L15 and their observations indexed per patient in chronological order (e.g. L09-2 is the second QACT of library patient 9), so that no date of service appears. Rows are sorted by forward rank. The two Dice columns are in different reference frames and must not be differenced or paired-tested; each locates one engine's own top-1 within its own scale.

| Patient | Pool | Scored by both | FlexiCT top-1 | MIND top-1 | Same | Dice, FlexiCT | Dice, MIND | MIND pick, FlexiCT rank | FlexiCT pick, MIND rank |
|---|---|---|---|---|---|---|---|---|---|
| P07 | 295 | 295 | D09-2 | D15 | No | 0.829 | 0.775 | 2 | 3 |
| P10 | 297 | 297 | D10-2 | D01 | No | 0.764 | 0.769 | 4 | 4 |
| P01 | 298 | 298 | D09-1 | D08 | No | 0.807 | 0.677 | 10 | 13 |
| P04 | 295 | 295 | D11 | D10-3 | No | 0.758 | 0.751 | 24 | 104 |
| P08 | 298 | 293 | D14 | D07 | No | 0.763 | 0.562 | 33 | 12 |
| P09 | 298 | 298 | D10-1 | D02-2 | No | 0.763 | 0.741 | 72 | 6 |
| P05 | 298 | 298 | D02-2 | D06 | No | 0.803 | 0.787 | 114 | 125 |
| P03 | 299 | 290 | D12 | D05 | No | 0.631 | 0.572 | 136 | 171 |
| P06 | 298 | 298 | D04 | D03 | No | 0.908 | 0.878 | 203 | 108 |
| P02 | 297 | 297 | D02-1 | D13 | No | 0.819 | 0.652 | 272 | 31 |